\documentclass[amsmath,amssymb,10pt,floatfix,A4,superscriptaddress,preprint]{revtex4}
\usepackage{graphicx,subfigure}
\usepackage{xcolor}
\usepackage{amssymb}
\usepackage[normalem]{ulem}
\usepackage{bm,tikz}
\usepackage{color}

\usepackage{graphicx,epsfig,epsf,color}
\usepackage{multirow}
\usepackage{float}
\usepackage[utf8x]{inputenc}
\usepackage{amssymb,amsmath,appendix}
\usepackage{slashed}
\usepackage{pict2e}
\usepackage{array}
\usepackage[colorlinks=true,citecolor=blue,urlcolor=blue,linkcolor=blue]{hyperref}

\begin{document}
\title{Bayesian analysis of density profile of light dark matter elucidating the properties of dark matter admixed neutron stars in the presence of hyperons}

\author{Debashree Sen}
\email{debashreesen88@gmail.com}
\affiliation{Department of Physics Education, Daegu University, Gyeongsan 38453, South Korea}

\author{Atanu Guha}
\email{atanu@cnu.ac.kr}
\affiliation{Department of Physics, Chungnam National University,\\ 99, Daehak-ro, Yuseong-gu, Daejeon-34134, South Korea}

\author{Chang Ho Hyun}
\email{hch@daegu.ac.kr}
\affiliation{Department of Physics Education, Daegu University, Gyeongsan 38453, South Korea}

\date{\today}




\begin{abstract}

We study the impact of symmetry energy ($S$), hyperons, and dark matter (DM) on structural and oscillatory properties of neutron stars (NSs). Uncertainty from hadronic equation of state for NSs is considered with 15 relativistic mean field models having slope parameter ($L_0$) of $S$ in range $40-120$ MeV. DM admixed NSs (DMANSs) are described with feeble interaction between light DM fermions ($\chi$) with hadronic matter in the presence of hyperons via scalar ($\eta$) and vector ($\xi$) dark mediators. The masses  $m_{\chi}$, $m_{\eta}$ and $m_{\xi}$ are related by self-interaction constraints from bullet cluster. DM self-interaction couplings are related to $m_{\chi}$ by relic density constraint. The DM density is taken as an exponential function of baryon density with a free parameter $\alpha$. Uncertainty from DM model is incorporated by exploring the dependence on $m_{\chi}$ and $\alpha$. Several DM search experiments have almost ruled out the existence of massive DM ($\gtrsim$ GeV). Lately, pursuit for sub-GeV DM has attracted significant attention. Therefore, we consider $m_{\chi}<$ 1 GeV and $\alpha \leq$ 0.1 such that the contribution of DM to the total mass of the DMANSs is $<10\%$. Comparing our results with various astrophysical constraints, we find that the HESS J1731-347 and GW170817 data are very important in determining the presence of light DM in NSs in moderate amount, relevant in the range $L_0\lesssim$ 58 MeV. Employing models of DMANSs that satisfy several observational data, we infer with Bayesian analysis, the likely ranges of $m_{\chi}$ and $\alpha$ are almost independent of the underlying hadronic model within 40 MeV $\lesssim$ $L_0$ $<$ 58 MeV. In the absence of DM and with the most probable values of $m_{\chi}$ and $\alpha$ obtained from the Bayesian inference, we calculate the frequencies of non-radial $f$- and $p_1$-modes oscillation of NSs/DMANSs.

\end{abstract}




\maketitle



\section{Introduction}
\label{sec:Introduction}

Neutron stars (NSs) are the densest objects in space from which we can detect signals in the form of electromagnetic waves, neutrinos, and gravitational waves (GWs). Because of its highest density in the visible universe, high compactness, and strong gravity, it could be a primary candidate in which dark matter (DM) is likely to reside and reveal its effects through gravitational interactions with particles that constitute the NS matter or, more generally, compact objects. Such stars are termed DM admixed NSs (DMANSs). However, DM is so elusive that an advanced understanding requires compilation of ample accurate data from various sources and events. Progress in the technique of astronomical observations opens new and deep insights into understanding the uncertainty in the state of matter at densities above the nuclear saturation, so it is timely and mandatory to explore the limit of our knowledge about the nature of dense nuclear matter and find ways to break through the present limit.

Equation of state (EoS) is a most fundamental quantity that determines the internal composition, structure, and bulk properties of the NSs. Because the mean density of a NS is $(2-3) \rho_0$ where $\rho_0$ is the nuclear saturation density, the uncertainty of EoS at high densities is critical to the internal and external properties of the star. There are a couple of sources that contribute to the uncertainty. In the nucleonic phase where only neutrons and protons constitute the matter, density dependence of the symmetry energy plays a crucial role to the uncertainty. At densities two or three times above the saturation density, new particles such as hyperons and $\Delta$'s can emerge \cite{Weissenborn:2011ut, Miyatsu:2013yta, Miyatsu:2025rzn, Sedrakian:2022ata, Huang:2022kej, ll2026, Parmar:2025csx, Marquez:2022gmu, Sen:2018tms, Sen:2018yyq, Sen:2021bms, Sen:2021cgl, Sen:2018qvo, Sen:2019kxt}, or new phases of matter like Bose-Einstein condensation of mesons \cite{Concepcion:2024duj} and/or transition to deconfined quark matter \cite{Annala:2019puf,Most:2018eaw,Sen:2018tms, Sen:2018yyq, Sen:2021bms, Sen:2021cgl, Sen:2024reu, Pal:2023quk, Sen:2022lig} can exist. EoS with these exotic degrees of freedom contain much larger uncertainties than those for the nucleonic matter. Therefore, to extract information from the NS data, one has to take the effect of accessible uncertainties into account, and derive physical results that are consistent with the range of allowed uncertainties.
 
In the 20th century, observation of NSs as pulsars was accessible in the binary systems, and only mass could be determined from the measurement. Most accurately measured masses are close to $1.4 M_\odot$, so this value became a canonical mass of the NS. In the new millennium, advancement in methodology and technology paved new ways for astronomical observation and they produced new results for the pulsar data. Most notable progresses are the determination of mass and radius from single objects and the detection of GWs from which we can extract the properties of the NSs other than mass and radius. These achievements indeed have great impacts on our understanding of the state of matter at high densities. Combination of mass-radius and tidal deformability from GW170817 data \cite{LIGOScientific:2018cki}, obtained by LIGO-Virgo collaboration, allows to constrain the EoS of canonical NSs more strongly than ever. Several recent measurements of the radius are obtained from PSR J0030+0451 \cite{Vinciguerra:2023qxq} and PSR J0437-4715 \cite{Choudhury:2024xbk} which have mass close to the canonical mass. In addition, recent simultaneous measurements of mass and radius of heavy-mass ($\gtrsim 2 M_\odot$) objects like PSR J0740+6620 \cite{Fonseca:2021wxt,Salmi:2024aum} by the NICER and small-mass ($\lesssim 1 M_\odot$) objects like HESS J1731-347 \cite{Doroshenko:2022} and PSR J1231-1411 \cite{Salmi:2024bss}, by HESS and NICER Collaborations, respectively, provide more stringent constraints to the EoS at low as well as high densities. Heavy-mass stars are relevant to the effect of exotic degrees of freedom like hyperons and quark matter. For small-mass stars, uncertainties are mostly relevant with the nucleonic EoS, so the data in a broad range $(1-2) M_\odot$ altogether provide constraints that link EoS from low to high densities.

Recently, an interesting issue has been raised about PSR J1231-1411. In the paper by Salmi {\it et al.} \cite{Salmi:2024bss}, mass and radius of the object are obtained from the NICER Collaboration data as
$1.04^{+0.05}_{-0.03} M_\odot$ and $12.6 \pm 0.3$ km, respectively. About four months later, an independent analysis on the same object was reported by Chinese group \cite{Qi:2025mpn}, in which mass is $1.12 \pm 0.07 M_\odot$ and radius is $9.01^{+0.88}_{-0.86}$ km. While the mass is similar in the two analyses, radii are divided and there is no overlapping region. If the object is neither a strange star nor a quark star, because the mass is about 1 $M_\odot$, it is likely that only neutrons and protons will consist of the core of the star. At present, further observational data and rigorous analyses are required for PSR J1231-1411 to obtain specific bound on its radius. In the nucleonic phase, most prominent uncertainty affecting the bulk properties of the NSs stems from the symmetry energy. Hyperons and deconfined quark phase can hardly play a role at this small mass star. However, there can be another source of uncertainty, DM.

Several pulsars are known for their fast rotation with high values of rotational frequency $\nu$. The fastest spinning pulsar currently known is PSR J1748-2446ad with $\nu$=716 Hz \cite{Hessels:2006ze}. Unfortunately, its mass and radius are not precisely determined. The second-fastest spinning pulsar known is PSR J0952-0607 ($\nu$=707.31 Hz) \cite{Bassa:2017zpe} with its accurately measured mass $\approx$ 2.35 $M_{\odot}$ \cite{Romani:2022jhd}. Taking into account the rotational effects on the structure of DMANSs, the measurements of rotational frequency and mass of PSR J0952-0607 can also constrain the EoS.

In this work, we study the structural and oscillation properties of DMANSs in the presence of hyperons. Few studies have been conducted in this context with many types of DM candidate \cite{Das:2021dru,Lopes:2024ixl, Mu:2023yqh, Shahrbaf:2025hsw,DelPopolo:2020pzh}. In this work, we consider feeble interaction between hadronic matter and light fermionic DM ($\chi$) whose mass does not exceed 1 GeV. The interaction is mediated by new physics scalar ($\phi$) and vector ($\xi$) mediators, following \cite{Sen:2021wev,Guha:2021njn,Guha:2024pnn}. The masses of the DM particles, $m_{\chi}$, $m_{\eta}$, and $m_{\xi}$ are consistently related to each other by following the self-interaction constraints from bullet cluster \cite{Randall:2007ph, Bradac:2006er, Tulin:2013teo, Tulin:2017ara, Hambye:2019tjt}. The couplings $y_{\eta}$ and $y_{\xi}$ associated with the self-interaction of DM are obtained by reproducing the observed non-baryonic relic density \cite{Belanger:2013oya, Gondolo:1990dk, Guha:2018mli}. 
In our earlier works \cite{Sen:2021wev,Guha:2021njn,Guha:2024pnn,Sen:2022pfr,Sen:2024yim,Jyothilakshmi:2024xtl} for DM admixed compact stars, we have considered a constant DM number density ($\rho_{\chi}$) following \cite{Panotopoulos:2017idn}. This assumption is still considered in many other recent works \cite{Lopes:2024ixl, Mu:2023yqh}. However, such an assumption of the constant $\rho_{\chi}$ may oversimplify the effects of DM in compact stars. Thus, it is suggested that the gravitational effects lead to a variable density profile of DM in compact stars \cite{Kumar:2025ytm}. Moreover, the assumption of $\rho_{\chi}$ (DM Fermi momentum $k_F^{\chi}$), being constant with the baryon density $\rho$ (radius) of the star, introduces thermodynamic inconsistency to the system \cite{Hajkarim_2025}. Due to gravitational interaction, there must be a radial variation of the DM chemical potential ($k_F^{\chi}$ and $\rho_{\chi}$) inside the star. Like all other baryonic particles, $\rho_{\chi}$ is therefore expected to be highest around the center of the star. The intense gravitational field of compact stars may accrete DM from its surroundings, and as the density of the core increases, the accreted DM tends to accumulate more towards the core. Therefore, $\rho_{\chi}$ is supposed to increase from the surface to the core. Studies, focusing on possible DM production in compact stars via dark decay of neutrons, also support this possibility \cite{Baym:2018ljz,Husain:2022brl}. Therefore, a novel feature of the present work is that we consider $\rho_{\chi}$ as an exponential function of the baryon density ($\rho$) of hadronic matter in terms of a free parameter $\alpha$. This makes $\rho_{\chi}$ variable with the radius of the star and helps us surpass both the issues of oversimplification and thermodynamic inconsistency. Such a form to connect $\rho_{\chi}$ and $\rho$ has not been considered before in the literature. Among many unknown properties associated with DM, we restrict the investigation to the effect of uncertainties originating from $m_{\chi}$ and $\rho_{\chi}$ via $\alpha$. Overall results from several studies have restricted the contribution of DM ($f$) to the total mass of DMANSs to a maximum of $\approx$10\% \cite{Zhang:2025rur,Sagun:2021oml,Karkevandi:2024vov,Deliyergiyev:2019vti} within the single-fluid approach for interaction of DM with baryonic matter. In this work, we restrict $\alpha$ in such a way that $f$ does not exceed 10\%. In the last decade, direct detection, indirect detection, and colliders are being operated as various search avenues for DM. Of them, several direct detection experiments such as LZ \cite{LZ:2022lsv}, XENON \cite{XENON:2019rxp}, DarkSide \cite{DarkSide:2018bpj}, CRESST \cite{CRESST:2019jnq}, and the collider experiment LHC \cite{Roy:2024ear} have already ruled out most of the parameter space (effective field theory expansion scale vs mass of DM) in the heavy DM mass regime ($m_{\chi} \geq 1~\rm{GeV}$). Therefore, low-mass DM of sub-GeV order is of current interest in recent times \cite{LZ:2025iaw, PandaX-II:2021kai, LUX:2018akb, Kim:2023onk, Guha:2024mjr, Kim:2024vxg, Leane:2025efj, Dutta:2024kuj, Lee:2024wzd, Choi:2024ism, Kim:2024ltz, Alhazmi:2025nvt, Mishra:2025juk, BetancourtKamenetskaia:2025rwk, Guha:2025tjf}. Thus, in the present work, we consider light DM $m_{\chi} <$ 1 GeV. 

In the hadronic sector, role of the symmetry energy ($S$) is taken into account by adopting 15 relativistic mean field (RMF) models with slope parameter ($L_0$) in $S$. $L_0$ is an important quantity whose precise value has not been determined well till date. The Lead Radius EXperiment-II (PREX-2) and the $^{48}$Ca Radius EXperiment (CREX) emerged independently with contrasting results as $L_0$ = (106 $\pm$ 37) MeV \cite{PREX:2021umo} and $L_0$=0--51 MeV \cite{CREX:2022kgg}, respectively, leading to the well-known PREX-CREX dilemma. However, \cite{Lattimer:2023rpe} showed that nuclear interactions imply $L_0$=53 $\pm$ 13 MeV, optimally satisfying both PREX and CREX measurements. This is also close to the range suggested by either nuclear mass measurements or neutron matter theory and is also consistent with nuclear dipole polarizability measurements. Within this range, the obtained values of radius and tidal deformability of a 1.4 $M_{\odot}$ NS are also consistent with the findings of NICER X-ray and LIGO/Virgo GW observational data \cite{Lattimer:2023rpe}. In the present work, we consider $L_0$ in the range $L_0=40-120$ MeV. Calibration of the EoS to the data of PSR J1231-1411 and HESS J1731-347 propagates its effect to the behavior of bulk properties in the high-mass region. To account for the uncertainties at high densities, we consider the hyperons in the baryon octet which are populated when the $\beta$-equilibrium conditions are fulfilled with neutrons and protons. SU(3) coupling scheme is adopted to fix the coupling constants in the interactions of hyperons following \cite{Weissenborn:2011ut, Miyatsu:2013yta, Miyatsu:2025rzn}. This is a more general and realistic scheme which can explain the heavy mass of strange quark. Also, it can distinguish between different hyperon-nucleon interaction potentials for various hyperon species. Extrapolation to high densities allows us to investigate the compatibility of the EoS determined by low-mass stars with the existence of $2 M_\odot$ stars and resolution of hyperon puzzle. Within the range of uncertainties stemming from the symmetry energy and DM, first, we calculate the mass-radius relation and tidal deformability of NSs in the presence and absence of hyperons, and then those for the DMANSs in both static and rotating conditions. The results, in each case, are compared with the astrophysical data obtained from pulsar observations and GW170817 detection. Performing Bayesian inference analysis, we derive the distribution of $m_{\chi}$ and $\alpha$ of DM that are consistent with the observed data. Finally, by using the most probable values inferred from the Bayesian analysis, we calculate the fundamental ($f$)-mode and first pressure ($p_1$)-mode frequencies of the non-radial oscillations of the DMANSs in full general relativistic treatment following \cite{1967ApJ...149..591T, Thorne:1969rba,Lindblom:1983ps, Detweiler:1985zz,Lu:2011zzd,Guha:2025ssq} and compare them with the results for hadronic NSs in presence of hyperons but in absence of DM. Stiff EoS tends to reduce the oscillation frequencies. Such a relationship has been intensively studied in the stars with various compositions \cite{Shahrbaf:2025hsw,Sen:2024yim,Guha:2024gfe, Kumar:2025cro,Sotani:2025hzb, Rather:2024mtd}. In the present work, we focus on whether softening or stiffening of the EoS by DM can have effects on the non-radial oscillation spectra that could be distinguished from the effects by other uncertainties and discuss the implication of the result to the future GW observations by upcoming GW detectors like the Advanced LIGO, A+, Cosmic Explorer, and Einstein Telescope etc.

The work is arranged in the following order. In Sec. \ref{Sec:Structure}, we discuss the structural properties of neutron stars when DM is admixed. Sec. \ref{sec:DMANS_Model} introduces the model for the baryons and DM. Sec. \ref{Sec:Bayesian analysis} details the process for the Bayesian analysis, and Sec. \ref{Sec:oscillation} provides the short comments on the non-radial oscillation. In the next section Sec. \ref{Results}, we present and discuss the results for the mass-radius relation, tidal deformability, Bayesian inference and the frequencies for the non-radial oscillation. The work is summarized in Sec. \ref{Conclusion}.

%


\section{Structural properties of dark matter admixed neutron stars}
\label{Sec:Structure}
The global properties like the gravitational mass ($M$) and the radius ($R$) of the stars in presence and absence of DM are computed in static and spherically symmetric conditions by integrating the Tolman-Oppenheimer-Volkoff (TOV) equations \cite{Tolman:1939jz,Oppenheimer:1939ne} using the corresponding EoS.
The tidal deformability is the quadrupole deformation of a star in a binary system due to the tidal field from its companion star. For an isolated star, the tidal deformability is calculated in the limit where the source of the static external quadrupolar tidal field is very far away \cite{Hinderer:2007mb}. We obtain the dimensionless tidal deformability ($\Lambda$) by following \cite{Hinderer:2007mb,Hinderer:2009ca}.

The structural properties like the mass and radius of the DMANSs in rotating condition are computed using the rotating neutron star (RNS) code \cite{Stergioulas:1994ea}. The mass and radius of the star are expected to increase in the rotating conditions compared to that in the static scenario because in the former case the Coriolis force comes into play.

The underlying model for formulating the EoS of DMANSs, required to compute the structural properties of the DMANSs, is discussed in the next section.

\section{Dark matter admixed neutron star model}
\label{sec:DMANS_Model}

We follow our previous works \cite{Sen:2021wev, Guha:2021njn,Guha:2024pnn} and in the core of NSs we invoke a feeble interaction of the dark fermion ($\chi$) with the hadronic matter ($\psi=N, Y$) through the scalar ($\eta$) and vector ($\xi$) dark mediators. The hadronic matter is composed of nucleons ($N=n,p$) and hyperons ($Y=\Lambda,\Sigma^{-,0,+},\Xi^{-,+}$) following SU(3) flavor symmetry for determining the vector meson couplings to the baryon octet. For the description of the hadronic matter we consider the density-dependent models DDLZ1 \cite{Wei:2020kfb}, DD2 \cite{Typel:2009sy}, DDME1 \cite{Niksic:2002yp}, DDME2 \cite{Lalazissis:2005de}, DDMEX \cite{Taninah:2019cku}, PKDD \cite{Long:2003dn} and models with non-linear self couplings like BigApple \cite{Fattoyev:2020cws}, NL3 \cite{Lalazissis:1996rd}, NL3$\omega \rho 1$ \cite{Pais:2016xiu}, NL3$\omega \rho 2$ \cite{Pais:2016xiu}, NL3$\omega \rho 3$ \cite{Pais:2016xiu}, NL3$\omega \rho 4$ \cite{Pais:2016xiu}, NL3$\omega \rho 5$ \cite{Pais:2016xiu}, and GM1 \cite{Glendenning:1991es}. Additionally, we also take the NL3$\omega \rho 6$ model. We consider those models that have moderate values of the slope parameter of symmetry energy ($L_0$) consistent with the results of the $^{48}$Ca Radius EXperiment (CREX) \cite{CREX:2022kgg} and the Lead Radius EXperiment-II (PREX-II) \cite{PREX:2021umo} experiments and also satisfy the maximum mass constraint obtained from the most massive pulsar PSR J0740+6620 \cite{Fonseca:2021wxt} in the presence of hyperons. The models with density independent, non-linear self couplings are called non-linear (NL) models in literature, and therefore we name the DM admixed NL models as DMNL models in this work. Similarly, the models with density-dependent couplings are called density-dependent (DD) models in the literature. Thus in this work, we name the DM admixed DD models as DMDD models. For the dark sector, a phenomenological treatment is considered to describe the self-interaction of non-relativistic DM with a Yukawa potential \cite{Tulin:2013teo,Guha:2024pnn}. We consider that $\eta$ and $\xi$ have their respective couplings as $y_{\eta}$ and $y_{\xi}$ with $\chi$ and we have the interaction Lagrangian $\mathcal{L}_{int}$ as
\begin{eqnarray} 
\mathcal{L}_{int} = \begin{cases}
y_\eta \eta \bar{\chi} \chi \\
y_\xi \bar{\chi} \gamma_\mu \chi \xi^\mu.
\end{cases} 
\end{eqnarray} 
The complete Lagrangian density for the DMNL models with SU(3) coupling scheme is given as
\begin{eqnarray} 
\mathcal{L}_{DMNL}&=&\sum_{B=N,Y} \bar{\psi}_B[\gamma_{\mu}(i\partial^{\mu} -g_{\omega B}\omega^{\mu} -g_{\rho B} \vec{\rho_\mu}\cdot \vec{\tau} -g_{\phi B}\phi^{\mu} -g_{\xi B}\xi^{\mu}) -(M_B -g_{\sigma B}\sigma -g_{\eta B}\eta)]\psi_B + \bar{\psi}_{\Lambda} g_{\sigma^* \Lambda} \sigma^* \psi_{\Lambda} \nonumber \\
&+& \frac{1}{2}\partial_{\mu}\sigma\partial^{\mu}\sigma + \frac{1}{2}\partial_{\mu}\sigma^*\partial^{\mu}\sigma^* -\frac{1}{2}m_{\sigma}^2{\sigma^2} -\frac{1}{2}m_{\sigma^*}^2{\sigma^*}^2 -\frac{1}{3}g_2\sigma^3 -\frac{1}{4}g_3\sigma^4  \nonumber\\
&-&\frac{1}{4}\omega_{\mu\nu}\omega^{\mu\nu} +\frac{1}{2}m_{\omega}^2\omega_{\mu}\omega^{\mu} +\frac{1}{4}c_3(\omega_{\mu}\omega^{\mu})^2 +\frac{1}{2}m_{\phi}^2{\phi^2} +\frac{1}{4}d_3(\phi_{\mu}\phi^{\mu})^2
-\frac{1}{4} \vec{R}_{\mu \nu} \cdot \vec{R}^{\mu \nu} +\frac{1}{2} m_\rho^2 \vec{\rho_\mu}\cdot \vec{\rho^\mu}  \nonumber\\
&+&\Lambda_{\omega \rho}(g_{\omega N}^2\omega^{\mu}\omega_{\mu})(g_{\rho N}^2\vec{\rho_\mu}\cdot \vec{\rho^\mu}) +\Lambda_{\phi \rho}(g_{\phi N}^2\phi^{\mu}\phi_{\mu})(g_{\rho N}^2\vec{\rho_\mu}\cdot \vec{\rho^\mu}) \nonumber\\
&+&\frac{1}{2} \partial_\mu \eta \partial^\mu \eta - \frac{1}{2}m_\eta^2 \eta^2 -\frac{1}{4} V_{\mu \nu} V^{\mu \nu} +\frac{1}{2} m_\xi^2 \xi_\mu \xi^\mu
+\bar{\chi} \left[\left( i \gamma_\mu \partial^\mu - y_\xi \gamma_\mu \xi^\mu \right) -\left(m_\chi + y_\eta \eta \right) \right] \chi ,
\label{eq:rmf_lagrangian_NL} 
\end{eqnarray} 
where, $\omega_{\mu\nu} = \partial_\mu \omega_\nu - \partial_\nu \omega_\mu$ and $\vec{R}_{\mu\nu} = \partial_\mu \vec{\rho_\nu} - \partial_\nu \vec{\rho_\mu}$, and $V_{\mu\nu} = \partial_\mu \xi_\nu - \partial_\nu \xi_\mu$.

For the DMDD models the Lagrangian density with SU(3) coupling scheme is given as
\begin{eqnarray} 
\mathcal{L}_{DMDD}&=&\sum_{B=N,Y} \bar{\psi}_B[\gamma_{\mu}(i\partial^{\mu} -g_{\omega B}(\rho)\omega^{\mu} -g_{\rho B}(\rho) \vec{\rho_\mu}\cdot \vec{\tau} -g_{\phi B}(\rho)\phi^{\mu} -g_{\xi B}\xi^{\mu}) -(M_B -g_{\sigma B}(\rho)\sigma -g_{\eta B}\eta)]\psi_B \nonumber \\ &+& \bar{\psi}_{\Lambda} g_{\sigma^* \Lambda} (\rho) \sigma^* \psi_{\Lambda}\nonumber \\
&+& \frac{1}{2}\partial_{\mu}\sigma\partial^{\mu}\sigma + \frac{1}{2}\partial_{\mu}\sigma^*\partial^{\mu}\sigma^* -\frac{1}{2}m_{\sigma}^2{\sigma^2} -\frac{1}{2}m_{\sigma^*}^2{\sigma^*}^2  \nonumber\\
&-&\frac{1}{4}\omega_{\mu\nu}\omega^{\mu\nu} +\frac{1}{2}m_{\omega}^2\omega_{\mu}\omega^{\mu} -\frac{1}{4} \vec{R}_{\mu \nu} \cdot \vec{R}^{\mu \nu} +\frac{1}{2}m_{\phi}^2{\phi^2} +\frac{1}{2} m_\rho^2 \vec{\rho_\mu}\cdot \vec{\rho^\mu} \nonumber\\
&+&\frac{1}{2} \partial_\mu \eta \partial^\mu \eta - \frac{1}{2}m_\eta^2 \eta^2 -\frac{1}{4} V_{\mu \nu} V^{\mu \nu} +\frac{1}{2} m_\xi^2 \xi_\mu \xi^\mu
+\bar{\chi} \left[\left( i \gamma_\mu \partial^\mu - y_\xi \gamma_\mu \xi^\mu \right) -\left(m_\chi + y_\eta \eta \right) \right] \chi .
\label{eq:rmf_lagrangian_DD} 
\end{eqnarray} 
In the pure hadronic sector the nucleons and hyperons interact via the scalar $\sigma$ meson, the vector $\omega$ and $\phi$ mesons, the strange vector meson $\sigma^*$ and the iso-vector $\rho$ meson. The total baryon density is 
\begin{eqnarray} 
\rho = \sum_{B=N,Y} \rho_B =\rho_N + \rho_Y =\frac{1}{2\pi^2} \sum_{B=N,Y} \gamma_B \int^{{k_F}_B}_0 dk_B k_B^2,  
\end{eqnarray}
${k_F}_B$ being the Fermi momenta of a particular baryon species $B$ and the spin degeneracy factor $\gamma_B$ is 2 in this case. The charge neutrality and chemical equilibrium conditions are imposed additionally in the NS matter in the presence of hyperons.

In the DMNL models, the mesons in the hadronic sector have density independent couplings $g_{\sigma B}$, $g_{\omega B}$, $g_{\phi B}$, $g_{\rho B}$ with the baryons $B$ (viz., nucleons ($N$) and hyperons ($Y$) while the $\sigma^*$ meson interacts only with the $\Lambda$ hyperon with coupling $g_{\sigma^* \Lambda}$. For the NL models BigApple, NL3, NL3$\omega \rho 1$, NL3$\omega \rho 2$, NL3$\omega \rho 3$, NL3$\omega \rho 4$, NL3$\omega \rho 5$, and GM1, $g_2$ and $g_3$ are the higher order scalar field coefficients while $c_3$ is the higher order vector field coefficient. These non-linear self couplings effectively account for the in-medium effects and in the BigApple, NL3$\omega \rho 1$, NL3$\omega \rho 2$, NL3$\omega \rho 3$, NL3$\omega \rho 4$, and NL3$\omega \rho 5$ models the coupling constant $\Lambda_{\omega \rho}$ modifies the density dependence of the symmetry energy \cite{Fattoyev:2020cws}. In the following Tab. \ref{tab:NL} we first show for the NL models, the meson couplings with nucleons ($g_{\sigma N}$, $g_{\omega N}$, $g_{\rho N}$, $g_2$, $g_3$, and $c_3$) in the SU(6) coupling scheme along with the mass of the mesons ($m_{\sigma}$, $m_{\omega}$ and $m_{\rho}$) and neutron ($m_n$) and proton ($m_p$) as adopted in the GM1 \cite{Glendenning:1991es}, BigApple \cite{Fattoyev:2020cws}, NL3 \cite{Lalazissis:1996rd}, NL3$\omega \rho 1$ \cite{Pais:2016xiu,Carriere:2002bx}, NL3$\omega \rho 2$ \cite{Pais:2016xiu,Carriere:2002bx}, NL3$\omega \rho 3$ \cite{Pais:2016xiu,Carriere:2002bx}, NL3$\omega \rho 4$ \cite{Pais:2016xiu,Carriere:2002bx}, and NL3$\omega \rho 5$ \cite{Pais:2016xiu,Carriere:2002bx} models according to the respective references. Additionally, we consider the NL3$\omega \rho 6$ model by fixing the value of $L_0$=58 MeV.
\begin{table}[!ht]
\caption{The density independent, non-linear meson-nucleon couplings and parameters adopted in the models GM1 \cite{Glendenning:1991es}, BigApple \cite{Fattoyev:2020cws}, NL3 \cite{Lalazissis:1996rd}, NL3$\omega \rho 1$ \cite{Pais:2016xiu,Carriere:2002bx}, NL3$\omega \rho 2$ \cite{Pais:2016xiu,Carriere:2002bx}, NL3$\omega \rho 3$ \cite{Pais:2016xiu,Carriere:2002bx}, NL3$\omega \rho 4$ \cite{Pais:2016xiu,Carriere:2002bx}, NL3$\omega \rho 5$ \cite{Pais:2016xiu,Carriere:2002bx}, and NL3$\omega \rho 6$ in SU(6) coupling scheme.}
\setlength{\tabcolsep}{4.0pt}
\begin{tabular}{ccccccccccccc}
\hline
\hline
Model & $m_n$ & $m_p$ & $m_{\sigma}$ & $m_{\omega}$ & $m_{\rho}$ & $g_{\sigma N}$ & $g_{\omega N}$ & $g_{\rho N}$ & $g_2$ & $g_3$ & $c_3$ & $\Lambda_{\omega \rho}$ \\
& (MeV) & (MeV) & (MeV) & (MeV) & (MeV) & & & & $({\rm fm}^{-1})$ &  \\
\hline
GM1 & 938 & 938 & 510 & 783 & 770 & 8.87443 & 10.60957 & 4.09772 & -9.7908 & -6.63661 & 0 & 0 \\
BigApple & 939 & 939 & 492.73 & 782.5 & 763 & 9.6699 & 12.316 & 14.1618 & -11.9214 & -31.6796 & 2.6843 & 0.0475 \\
NL3 & 939 & 939 & 508.1940 & 782.501 & 763 & 10.2169 & 12.8675 & 8.948 & -10.4707 & -28.8851 & 0 & 0 \\
NL3$\omega \rho 1$ & 939 & 939 & 508.1940 & 782.501 & 763 & 10.2169 & 12.8675 & 9.2141 & -10.4707 & -28.8851 & 0 & 0.005 \\
NL3$\omega \rho 2$ & 939 & 939 & 508.1940 & 782.501 & 763 & 10.2169 & 12.8675 & 9.5341 & -10.4707 & -28.8851 & 0 & 0.01 \\
NL3$\omega \rho 3$ & 939 & 939 & 508.1940 & 782.501 & 763 & 10.2169 & 12.8675 & 9.8944 & -10.4707 & -28.8851 & 0 & 0.015 \\
NL3$\omega \rho 4$ & 939 & 939 & 508.1940 & 782.501 & 763 & 10.2169 & 12.8675 & 10.2956 & -10.4707 & -28.8851 & 0 & 0.02 \\
NL3$\omega \rho 5$ & 939 & 939 & 508.1940 & 782.501 & 763 & 10.2169 & 12.8675 & 10.7517 & -10.4707 & -28.8851 & 0 & 0.025 \\
NL3$\omega \rho 6$ & 939 & 939 & 508.1940 & 782.501 & 763 & 10.2169 & 12.8675 & 11.0465 & -10.4707 & -28.8851 & 0 & 0.0273 \\
\hline
\hline
\end{tabular}
\label{tab:NL}
\end{table}

In models like DDLZ1 \cite{Wei:2020kfb}, DDMEX \cite{Taninah:2019cku}, DDME2  \cite{Lalazissis:2005de}, DD2 \cite{Typel:2009sy}, DDME1 \cite{Niksic:2002yp}, and PKDD \cite{Long:2003dn}, $g_2$=$g_3$=$c_3$=0 and the in-medium effects are treated with the density-dependent couplings following the Typel-Wolter ansatz \cite{Lu:2011wy}. For such models, the nucleon-meson couplings are given as
\begin{eqnarray}
g_{iN}(\rho)=g_{iN} a_i \frac{1+b_i(x+d_i)^2}{1+c_i(x+d_i)^2},
\end{eqnarray}
where $i=\sigma, \omega, \phi, \sigma^*$ and $x=\rho/\rho_0$ while
\begin{eqnarray}
g_{\rho N}(\rho)=g_{\rho N} \exp [a_{\rho}(x-1)].
\end{eqnarray}
The coefficients of the meson coupling constants $g_{iN}$ and $g_{\rho N}$ of the DDLZ1 model are the values at zero density ($\rho=0$), while for the other models, $g_{iN}$ and $g_{\rho N}$ are calculated at nuclear saturation density ($\rho=\rho_0$). In the following Tab. \ref{tab:DD} we display for the DD models, the meson couplings with nucleons ($g_{\sigma N}$, $g_{\omega N}$, and $g_{\rho N}$) in the SU(6) coupling scheme along with the mass of the mesons ($m_{\sigma}$, $m_{\omega}$ and $m_{\rho}$) and neutron ($m_n$) and proton ($m_p$) as adopted in the DDLZ1 \cite{Wei:2020kfb}, DDMEX \cite{Taninah:2019cku}, DDME2  \cite{Lalazissis:2005de}, DD2 \cite{Typel:2009sy}, DDME1 \cite{Niksic:2002yp}, and PKDD \cite{Long:2003dn} models according to the respective references. 
\begin{table}[h]
\caption{The density dependent meson-nucleon couplings and parameters adopted in the models DD2 \cite{Typel:2009sy}, DDME1 \cite{Niksic:2002yp}, DDME2  \cite{Lalazissis:2005de}, DDMEX \cite{Taninah:2019cku}, PKDD \cite{Long:2003dn}, and DDLZ1 \cite{Wei:2020kfb} in SU(6) coupling scheme.}
\setlength{\tabcolsep}{5.0pt}
\begin{tabular}{cccccccccccccccccc}
\hline
\hline
Model & $m_n$ & $m_p$ & $m_{\sigma}$ & $m_{\omega}$ & $m_{\rho}$ & $g_{\sigma N}$ & $g_{\omega N}$ & $g_{\rho N}$ \\
& (MeV) & (MeV) & (MeV) & (MeV) & (MeV) & & &\\
\hline
DD2 & 939.56536 & 938.27203 & 546.212459 & 783 & 763 & 10.686681 & 13.342362 & 7.25388 \\
DDME1 & 939 & 939 & 549.5255 & 783 & 763 & 10.4434 & 12.8939 & 7.6106 \\ 
DDME2 & 939 & 939 & 550.123 & 783 & 763 & 10.5396 & 13.0189 & 7.3672 \\ 
DDMEX & 939 & 939 & 547.3327 & 783 & 763 & 10.7067 & 13.3388 & 7.2380 \\ 
PKDD & 939.5731 & 938.2796 & 555.5112 & 783 & 763 & 10.7385 & 13.1476 & 8.5996 \\
DDLZ1 & 938.9 & 938.9 & 538.619216 & 783 & 763 & 12.001429 & 14.292525 & 7.575467 \\
\hline
Model & $a_{\sigma}$ & $b_{\sigma}$ & $c_{\sigma}$ & $d_{\sigma}$ & $a_{\omega}$ & $b_{\omega}$ & $c_{\omega}$ & $d_{\omega}$ & $a_{\rho}$ \\
\hline
DD2 & 1.357630 & 0.634442 & 1.005358 & 0.575810 & 1.369718 & 0.496475 & 0.817753 & 0.638452 & 0.983955 \\
DDME1 & 1.3854 & 0.9781 & 1.5342 & 0.4661 & 1.3879 & 0.8525 & 1.3566 & 0.4957 & 0.5008 \\
DDME2 & 1.3881 & 1.0943 & 1.7057 & 0.4421 & 1.3892 & 0.9240 & 1.4620 & 0.4775 & 0.5647 \\
DDMEX & 1.3970 & 1.3350 & 2.0671 & 0.4016 & 1.3936 & 1.0191 & 1.6060 & 0.4556 & 0.6202 \\
PKDD & 1.327423 & 0.435126 & 0.691666 & 0.694210 & 1.342170 & 0.371167 & 0.611397 & 0.738376 & 0.183305 \\
DDLZ1 & 1.062748 & 1.763627 & 2.308928 & 0.379957 & 1.059181 & 0.418273 & 0.538663 & 0.786649 & 0.776095 \\
\hline
\hline
\end{tabular}
\label{tab:DD}
\end{table}

In the symmetric nuclear matter (SNM) the nuclear saturation properties like the saturation density ($\rho_0$), binding energy per nucleon ($B/A$), nuclear incompressibility ($K_0$), skewness coefficient ($Q_0$), symmetry energy ($J_0$), slope ($L_0$), and the curvature parameter ($K_{sym}$) of the nuclear symmetry energy of the NL and DD models can be found in their respective references and are also listed in works like \cite{Xia:2022dvw, Huang:2022kej}.

With respect to the strange mesons, we consider $m_{\sigma^*}$=980 MeV and $m_{\phi}=1020$ MeV for all the models following \cite{Huang:2022kej}. The masses of the hyperons are taken as $m_{\Lambda}$=1115.68 MeV, $m_{\Sigma^{-,0,+}}$=(1197.45, 1192.64, 1189.37) MeV and $m_{\Xi^{-,0}}$=(1321.71, 1314.86) MeV in all the models \cite{Huang:2022kej}. We consider the SU(3) coupling scheme to obtain the baryon-meson couplings following \cite{Weissenborn:2011ut, Miyatsu:2013yta, Miyatsu:2025rzn}. In this scheme, the mixing between the $\omega$ and $\phi$ mesons is as follows
\begin{eqnarray}
g_{\omega B} = g_1 \cos\theta_V + g_8 c_B \sin\theta_V 
\end{eqnarray}
and
\begin{eqnarray}
g_{\phi B} = -g_1 \sin\theta_V + g_8 c_B \cos\theta_V .
\end{eqnarray}
Here $c_B$ is calculated for each baryon species via $\alpha_V=\frac{F}{F+D}$, where $F$ and $D$ are the anti-symmetric and symmetric terms of the interaction
of the meson octet and the baryons \cite{deSwart:1963pdg,Weissenborn:2011ut, Miyatsu:2013yta, Miyatsu:2025rzn}. In the present work, we fix the mixing angle $\theta_V$=35.26$^{\circ}$, $Z_V=\frac{g_8}{g_1}$=0.2, and $\alpha_V$=1 because we aim to study the effects of DM contribution explicitly. Due to the non-ideal $\omega-\phi$ mixing in the SU(3) scheme, in all the DD models and NL models that do not have higher order terms of $\omega$ meson and the mixing terms between the vector mesons (i.e., only the GM1 model), $g_{\omega N}$ is now modified as
\begin{eqnarray}
g_{\omega N}=\frac{\tilde{g}_{\omega N}}{\sqrt{1+R_{\phi N}^2\big(\frac{m_{\omega}}{m_{\phi}}\big)^2}}
\end{eqnarray}
where $\tilde{g}_{\omega N}$ is the $\omega-N$ coupling in the ideal mixing (SU(6) coupling scheme) case. This modified value must satisfy all the saturation properties in SNM. As a result of the non-ideal mixing in the SU(3) scheme, the $g_{\phi N}$ coupling is no longer zero, but 
\begin{eqnarray}
g_{\phi N}=-\frac{\tan\theta_V + \frac{1}{\sqrt{3}}Z_V(1-4\alpha_V)}{1-\frac{1}{\sqrt{3}}Z_V(1-4\alpha_V)\tan\theta_V}g_{\omega N} .
\end{eqnarray}
In the NL3$\omega \rho$1, NL3$\omega \rho$2, NL3$\omega \rho$3, NL3$\omega \rho$4, NL3$\omega \rho$5, and NL3$\omega \rho 6$ models, since the value of $L_0$ changes from the original NL3 model due to the introduction of the mixing terms between the vector mesons, the values of $g_{\rho N}$ and $\Lambda_{\omega \rho}$ are modified. The couplings are re-adjusted by preserving the value of symmetry energy at $\rho$=0.11 fm$^{-3}$. For models containing the higher order terms of $\omega$ meson and the mixing terms between the vector mesons, i.e., the BigApple, NL3$\omega \rho$1, NL3$\omega \rho$2, NL3$\omega \rho$3, NL3$\omega \rho$4, NL3$\omega \rho$5, and NL3$\omega \rho 6$ models, the values of $g_{\omega N}$, $c_3$, and $\Lambda_{\omega \rho}$ need to be recalculated and two additional couplings viz, $d_3$ and $\Lambda_{\phi \rho}$ need to be introduced, as seen from Eq. (\ref{eq:rmf_lagrangian_NL}) in the SU(3) coupling scheme. The couplings are calculated at $\rho_0$ by preserving the nuclear saturation properties $B/A(\rho_0)$, $J_0$, $L_0$, $K_0$, and $K_{sym}$ of the particular model in SNM \cite{Miyatsu:2025rzn}. In Tab. \ref{tab:model_SU3} the modified meson-nucleon couplings ($g_{\omega N}$, $g_{\rho N}$, $c_3$, and $\Lambda_{\omega \rho}$) and additional couplings ($d_3$ and $\Lambda_{\phi \rho}$) for SU(3) coupling scheme are shown.
\begin{table}[!ht]
\caption{The modified meson-nucleon couplings and additional couplings for SU(3) coupling scheme.}
\setlength{\tabcolsep}{10.0pt}
\begin{tabular}{ccccccccccccc}
\hline
\hline
Model & $g_{\omega N}$ & $g_{\rho N}$ & $c_3$ & $d_3$ & $\Lambda_{\omega \rho}$ & $\Lambda_{\phi \rho}$ \\
\hline
BigApple & 12.023328 & - & 2.8506  & 2.3742 & 0.05203 & 0.09838 \\
NL3$\omega \rho 1$ & 12.56199 & 9.243737 & 9.05$\times10^{-6}$ & -0.012698 & 0.0055 & 0.0047 & \\
NL3$\omega \rho 2$ & 12.56199 & 9.5708 & 3.39$\times10^{-6}$ & -0.02232 & 0.01098 & 0.01003 \\
NL3$\omega \rho 3$ & 12.56199 & 9.93541 & 3.22$\times10^{-5}$ & -0.02882 & 0.01647 & 0.01526 \\
NL3$\omega \rho 4$ & 12.56199 & 10.3450 & 2.24$\times10^{-6}$ & -0.03229 & 0.02197 & 0.0131 \\
NL3$\omega \rho 5$ & 12.56199 & 10.8099 & 7.05$\times10^{-6}$ & -0.03441 & 0.02747 & 0.0227 \\
NL3$\omega \rho 6$ & 12.56199 & 11.04655 & 2.88$\times10^{-6}$ & -0.03472 & 0.03 & 0.02262 \\
NL3 & 12.56199 & - & - & - & - & - \\
GM1 & 10.356696 &- &-  &-  &-  &- \\
DD2   & 13.02435 &- &-  &-  &-  &- \\
DDME1 & 12.58658 &- &-  &-  &-  &- \\ 
DDME2 & 12.70860 &- &-  &-  &-  &- \\  
DDMEX & 13.02087 &- &-  &-  &-  &- \\ 
PKDD  & 12.83423  &-&-  &-  &-  &- \\
DDLZ1 & 14.2925 &- &-  &-  &-  &- \\
\hline
\hline
\end{tabular}
\label{tab:model_SU3}
\end{table}

We define the hyperon-meson couplings as $R_{iY}=g_{iY}/g_{iN}$, where $i=\sigma, \omega, \rho, \phi$ and $\sigma^*$. The couplings $R_{\omega Y}$ and $R_{\phi Y}$ can be obtained in terms of $z_V$ and $\alpha_V$ while $R_{\rho \Lambda}$=0 and $R_{\rho \Sigma}$ and $R_{\rho \Xi}$ are obtained in terms of only $\alpha_V$ \cite{Weissenborn:2011ut, Miyatsu:2025rzn}. Once we obtain the couplings $R_{\omega Y}$, the couplings $R_{\sigma Y}$ are obtained using the potential depths ($U_Y^{SNM} (\rho_0)$) of the corresponding hyperon species in SNM at $\rho_0$. In this work, we take $U_\Lambda^{SNM}(\rho_0)$=-30 MeV, $U_\Sigma^{SNM}(\rho_0)$=+30 MeV and $U_\Xi^{SNM}(\rho_0)$=-14 MeV following \cite{Huang:2022kej}. The coupling $R_{\sigma^* \Lambda}$ is obtained using the potential depth ($U_\Lambda^{\Lambda}(\rho_0)$) of the $\Lambda$ hyperon in pure $\Lambda$ matter at saturation density in the presence of the $\phi$ meson. In the present work, we consider $U_\Lambda^{\Lambda}(\rho_0)$=-10 MeV following \cite{Huang:2022kej}.

The vacuum expectation values (VEVs) or the equation of motion (EoM) of the baryons and mesonic mediator fields in the RMF approximation are modified by DM. For the DD models, they are as follows.\\
For the baryons
\begin{eqnarray}
\sum_{B=N,Y}\left[\left(g_{\omega B}(\rho)\omega_0 +\frac{g_{\rho B}(\rho)}{2}\tau_{3B}\rho_{03} +g_{\phi B}(\rho)\phi_0 + g_{\xi B}\xi_0 + \Sigma^R\right) -m_B^{\star} \right]\psi_B = 0.   
\label{eq:psi0DD}
\end{eqnarray}
Here, the rearrangement term $\Sigma^R$ in the presence of hyperons is given by \cite{Lenske:1995wyj} as
\begin{eqnarray} 
\Sigma^R = \sum_{B=N,Y}-\frac{dg_{\sigma B}(\rho)}{d\rho}\sigma_0\rho_{S B} -\frac{dg_{\sigma^* \Lambda}(\rho)}{d\rho}\sigma^*_0\rho_{S \Lambda} + \frac{dg_{\omega B}(\rho)}{d\rho}\omega_0\rho_B + \frac{1}{2}\frac{dg_{\rho B}(\rho)}{d\rho}\tau_{3B}\rho_{03}\rho_B + \frac{dg_{\phi B}(\rho)}{d\rho}\phi_0\rho_B.
\end{eqnarray} 
Here, $B=N, Y$, $\tau_{3B}/2$ is the third component of isospin and $\rho_{SB}$ is the scalar density of individual baryon given by
\begin{eqnarray} 
\rho_{SB}=\frac{\gamma_B}{2\pi^2}\int_0^{{k_F}_B}dk k_B^2 \frac{m_B^{\star}}{\sqrt{k_B^2+{m_B^{\star}}^2}} .
\end{eqnarray} 
For the mesons
\begin{eqnarray}
m_{\sigma}^2\sigma_0 = \sum_{B=N,Y} g_{\sigma B}(\rho) \rho_{SB} ,
\label{eq:sigma0DD}
\end{eqnarray}
\begin{eqnarray}
m_{\omega}^2\omega_0 = \sum_{B=N,Y} g_{\omega B}(\rho) \rho_B ,
\label{eq:omega0DD}
\end{eqnarray}
\begin{eqnarray}
m_{\rho}^2\rho_{03} = \sum_{B=N,Y} \frac{g_{\rho B}(\rho)}{2}\tau_{3B} \rho_B ,
\label{eq:rho0DD}
\end{eqnarray}
\begin{eqnarray}
m_{\phi}^2\phi_0 = \sum_{B=N,Y} g_{\phi B}(\rho) \rho_B ,
\label{eq:phi0DD}
\end{eqnarray}
and
\begin{eqnarray}
m^2_{\sigma^*}\sigma_0^* = \sum_{B=N,Y} g_{\sigma^* \Lambda}(\rho) \rho_{S \Lambda} .
\label{eq:sigma_str0DD}
\end{eqnarray}
For the NL models, the EoM of the mesons are modified as below,\\
\begin{eqnarray}
m_{\sigma}^2\sigma_0 +g_2\sigma_0^2 +g_3\sigma_0^3 = \sum_{B=N,Y} g_{\sigma B}\rho_{SB} ,
\label{eq:sigma0NL}
\end{eqnarray}
\begin{eqnarray}
m_{\omega}^2\omega_0 +c_3\omega_0^3 +2\Lambda_{\omega \rho}(g_{\omega N}^2\omega_0)(g_{\rho N}^2{\rho_{03}^2}) = \sum_{B=N,Y} g_{\omega B} \rho_B ,
\label{eq:omega0NL}
\end{eqnarray}
\begin{eqnarray}
m_{\rho}^2\rho_{03} +2\Lambda_{\omega \rho}(g_{\omega N}^2\omega_0^2)(g_{\rho N}^2{\rho_{03}}) = \sum_{B=N,Y} \frac{g_{\rho B}}{2}\tau_{3B} \rho_B ,
\label{eq:rho0NL}
\end{eqnarray}
and
\begin{eqnarray}
m_{\phi}^2\phi_0 +d_3\phi_0^3 +2\Lambda_{\phi \rho} (g_{\phi N}^2\phi_0^2)(g_{\rho N}^2\rho_{03}^2) = \sum_{B=N,Y} g_{\phi B} \rho_B .
\label{eq:phi0NL}
\end{eqnarray}
The EoM of the baryons $\psi_B$ and $\sigma^*$ mesons for the NL models remain the same as those for the DD models given by Eqs. (\ref{eq:psi0DD}) and (\ref{eq:sigma_str0DD}), respectively, except for the fact that the couplings are density independent and $\Sigma^R$=0 in Eq. (\ref{eq:psi0DD}).

The dark bosons $\eta$ and $\xi$ interact with the baryons $\psi_B$ with a very feeble coupling strength. In case of nucleons, we consider $g_{\eta N}=g_{\xi N}\sim$10$^{-3}$ assuming that both the scalar and vector channels contribute equally to the relic abundance \cite{Sen:2021wev, Guha:2021njn,Guha:2024pnn}. Since we consider charge neutral DM, the DM admixed NS matter remains charge neutral. The VEVs or the EoM of the DM mediator fields in RMF approximation are
\begin{eqnarray} 
\eta_0=\frac{m^\star_\chi-m_\chi}{y_{\eta}},
\end{eqnarray} 
and
\begin{eqnarray} 
\xi_0 = \frac{1}{m_\xi^2}\left(\sum_{B=N,Y} g_{\xi B} \rho_B + y_\xi \rho_\chi\right).
\end{eqnarray} 
In the present work, we assume that the fermionic DM density and the baryon density are related by the following form as considered in case of DMSQSs in \cite{Sen:2025ndg}
\begin{eqnarray}
\rho_{\chi}= \rho_{sc} \alpha(e^{\rho/\rho_{sc}} - 1),
\label{eq:rho_chi}
\end{eqnarray}
where, $\rho_{sc}$ is the reference number density. $\alpha$ controls the behavior and distribution of DM relative to the baryonic matter within the star. In the present work we aim to constrain the values of $\alpha$, $\rho_{sc}$, and $m_{\chi}$ in the light of the different bounds on the DM parameters and the astrophysical constraints on the structural properties of the compact stars. The importance of this particular form of the density profile of DM in Eq. (\ref{eq:rho_chi}) is that it includes the contributions from the different powers of $\rho$. Therefore, the DM Fermi momentum ${k_F}_{\chi}$ is no longer constant throughout the density range of the star and can be written as
\begin{eqnarray}
k_F^{\chi}={\left(\frac{6\pi^2}{\gamma_{\chi}}\rho_{\chi}\right)}^{1/3},
\end{eqnarray}
where $\gamma_{\chi}$=2. In case of the DD models, the modified effective mass of each baryon due to DM interaction with the baryons is 
\begin{eqnarray} 
m_B^{\star}=M_B -g_{\sigma B}(\rho)\sigma_0 -g_{\sigma^* B}(\rho)\sigma^*_0 -g_{\eta B}\eta_0
\label{eq:mb}
\end{eqnarray} 
while the modified chemical potential of each baryon is
\begin{eqnarray} 
\mu_B=\sqrt{k_B^2 + {m_B^{\star}}^2} + g_{\omega B}(\rho)\omega_0 + \frac{g_{\rho B}(\rho)}{2}\tau_{3B}\rho_{03} + g_{\phi B}(\rho)\phi_0 +\Sigma^R + g_{\xi B}\xi_0 .
\label{eq:muB}
\end{eqnarray} 
For the NL models, $\Sigma^R$=0 in Eq. (\ref{eq:muB}) and the couplings are density independent in both Eqs. (\ref{eq:mb}) and (\ref{eq:muB}). 

The conditions for the appearance of hyperons in the $\beta$-equilibrated NS matter, in the interacting scenario, are also modified in the presence of DM in the following way. The hyperons appear when $\mu_n + Q_Y\mu_e \geq \mu_Y(k_Y=0)$, where $\mu_Y(k_Y=0)$ is the chemical potential of individual hyperon at rest and $Q_Y$ is the charge of each hyperon. The rest mass of the hyperons is given by
\begin{eqnarray}
m^{\star}_Y(k_Y=0)=m_Y - R_{\sigma Y}g_{\sigma N}(\rho)\sigma_0 - R_{\eta Y}g_{\eta}\eta_0 .
\label{eq:mY*}
\end{eqnarray}
The hyperon chemical potentials at rest are given as below.\\
For the $\Lambda$ and $\Sigma^0$ hyperons
\begin{eqnarray}
\mu_Y(k_Y=0)=m^\star_Y(k_Y=0) + R_{\omega Y}g_{\omega N}(\rho) \omega_0 + R_{\phi Y}g_{\omega N}(\rho) \phi_0 + \Sigma^R  + R_{\xi Y}g_{\xi}\xi_0,  
\label{eq:muLamSig0}
\end{eqnarray}
for the $\Sigma^{-,+}$ hyperon
\begin{eqnarray}
\mu_{\Sigma^{-,+}}(k_{\Sigma^{-,+}}=0)=m^\star_{\Sigma}(k_{\Sigma}=0) + R_{\omega \Sigma}g_{\omega N}(\rho) \omega_0 + R_{\phi \Sigma}g_{\omega N}(\rho) \phi_0 \mp \frac{1}{2}R_{\rho \Sigma}g_{\rho N}(\rho)\rho_{03} + \Sigma^R  + R_{\xi Y}g_{\xi}\xi_0,
\label{eq:muSig-+}
\end{eqnarray}
and for the $\Xi^{-,0}$ hyperon
\begin{eqnarray}
\mu_{\Xi^{-,0}}(k_{\Xi^{-,0}}=0)=m^\star_{\Xi}(k_{\Xi}=0) + R_{\omega \Xi}g_{\omega N}(\rho) \omega_0 + R_{\phi \Xi}g_{\omega N}(\rho) \phi_0 \mp \frac{1}{2}R_{\rho \Xi}g_{\rho N}(\rho)\rho_{03} + \Sigma^R  + R_{\xi Y}g_{\xi}\xi_0 .
\label{eq:muCas}
\end{eqnarray}
The couplings in Eqs. (\ref{eq:mY*}), (\ref{eq:muLamSig0}), (\ref{eq:muSig-+}), and (\ref{eq:muCas}) are density independent for the NL models GM1 and BigApple. We consider the DM coupling with hyperons as $R_{\eta Y}(=g_{\eta Y}/g_{\eta N})=R_{\xi Y}(=g_{\xi Y}/g_{\xi N})=$ 0.5. We have checked that the variation of this coupling strength does not bring any significant change in the EoS and structural properties of the DMANSs.

The complete expressions for the EoS are also modified as a result of the presence of DM. The energy density ($\varepsilon_{DMDD}$) for the DD models is given as
\begin{eqnarray}
\varepsilon_{DMDD}=\frac{1}{2}m_{\sigma}^2\sigma_0^2 + \frac{1}{2}m_{\sigma^*}^2{\sigma^*_0}^2 -\frac{1}{2}m_{\omega}^2\omega_0^2 -\frac{1}{2}m_{\rho}^2\rho_{03}^2 -\frac{1}{2}m_{\phi}^2\phi_0^2 \nonumber \\
+\sum_{B=N,Y} g_{\omega B}(\rho)\omega_0\rho_B +\sum_{B=N,Y} \frac{g_{\rho B}(\rho)}{2}\tau_{3B}\rho_{03}\rho_B +\sum_{B=N,Y} g_{\phi B}(\rho)\phi_0\rho_B \nonumber \\
+ \sum_{B=N,Y} \frac{\gamma_B}{2\pi^2} \int_0^{{k_F}_B} \sqrt{k_B^2 + {m_B^{\star}}^2}~k_B^2 dk + \sum_{l=e,\mu} \frac{\gamma_l}{2 \pi^2} \int_0^{k_l} \sqrt{k_l^2 + m_l^2}~ k_l^2 dk_l \nonumber \\
+\frac{1}{2}m_{\eta}^2\eta_0^2 +\sum_{B=N,Y} g_{\xi B}\xi_0\rho_B +y_{\xi}\xi_0\rho_{\chi} -\frac{1}{2}m_{\xi}^2\xi_0^2 +\frac{\gamma_{\chi}}{2\pi^2} \int_0^{k_F^{\chi}} \sqrt{k_{\chi}^2 + {m_{\chi}^{\star}}^2}~ k_{\chi}^2 dk{\chi} .
\label{eDD}
\end{eqnarray} 
In case of NL models, the energy density ($\varepsilon_{DMNL}$) is given by
\begin{eqnarray}
\varepsilon_{DMNL}&=& \frac{1}{2}m_{\sigma^*}^2{\sigma^*_0}^2 +\frac{1}{2}m_{\sigma}^2\sigma_0^2 +\frac{1}{3}g_2\sigma_0^3 + \frac{1}{4}g_3\sigma_0^4 -\frac{1}{2}m_{\omega}^2\omega_0^2 -\frac{1}{4}c_3\omega_0^4 -\frac{1}{4}d_3\phi_0^4 -\frac{1}{2}m_{\rho}^2\rho_{03}^2 -\frac{1}{2}m_{\phi}^2\phi_0^2\nonumber \\ 
&-&\Lambda_{\omega \rho}(g_{\omega N}^2\omega_0^2)(g_{\rho N}^2{\rho_{03}^2}) -\Lambda_{\phi \rho}(g_{\phi N}^2\omega_0^2)(g_{\rho N}^2{\rho_{03}^2}) \nonumber \\ 
&+&\sum_{B=N,Y} g_{\omega B}\omega_0\rho_B +\sum_{B=N,Y} \frac{g_{\rho B}}{2}\tau_{3B}\rho_{03}\rho_B +\sum_{B=N,Y} g_{\phi B}\phi_0\rho_B \nonumber \\
&+& \sum_{B=N,Y} \frac{\gamma_B}{2\pi^2} \int_0^{{k_F}_B} \sqrt{k_B^2 + {m_B^{\star}}^2}~k_B^2 dk + \sum_{l=e,\mu} \frac{\gamma_l}{2 \pi^2} \int_0^{k_l} \sqrt{k_l^2 + m_l^2}~ k_l^2 dk_l \nonumber \\ &+&\frac{1}{2}m_{\eta}^2\eta_0^2 +\sum_{B=N,Y} g_{\xi B}\xi_0\rho_B +y_{\xi}\xi_0\rho_{\chi} -\frac{1}{2}m_{\xi}^2\xi_0^2 +\frac{\gamma_{\chi}}{2\pi^2} \int_0^{k_F^{\chi}} \sqrt{k_{\chi}^2 + {m_{\chi}^{\star}}^2}~ k_{\chi}^2 dk{\chi} .
\label{eNL}
\end{eqnarray} 
The pressure ($P_{DMDD}$) for the DD models is given as
\begin{eqnarray}
P_{DMDD}=-\frac{1}{2}m_{\sigma}^2\sigma_0^2 - \frac{1}{2}m_{\sigma^*}^2{\sigma^*_0}^2 +\frac{1}{2}m_{\omega}^2\omega_0^2 +\frac{1}{2}m_{\rho}^2\rho_{03}^2 +\frac{1}{2}m_{\phi}^2\phi_0^2 \nonumber \\
+ \sum_{B=N,Y} \frac{\gamma_B}{6\pi^2} \int_0^{{k_F}_B} \frac{k_B^4 dk}{\sqrt{k_B^2 + {m_B^{\star}}^2}} + \sum_{l=e,\mu} \frac{\gamma_l}{6 \pi^2} \int_0^{k_l} \frac{k_l^4 dk_l}{\sqrt{k_l^2 + m_l^2}} \nonumber \\ 
-\frac{1}{2}m_{\eta}^2\eta_0^2 +\frac{1}{2}m_{\xi}^2\xi_0^2 +\frac{\gamma_\chi}{6 \pi^2} \int_0^{k_F^{\chi}} \frac{k_\chi^4 dk_\chi}{\sqrt{k_\chi^2 + {m_\chi^{\star}}^2}} + \rho\Sigma^R
\label{pDD}
\end{eqnarray} 
and for the NL models, the pressure ($P_{DMNL}$) is
\begin{eqnarray}
P_{DMNL}=&-&\frac{1}{2}m_{\sigma^*}^2{\sigma^*_0}^2 -\frac{1}{2} m_{\sigma}^2\sigma_0^2 -\frac{1}{3}g_2\sigma_0^3 -\frac{1}{4}g_3\sigma_0^4 +\frac{1}{2}m_{\omega}^2\omega_0^2 +\frac{1}{4}c_3\omega_0^4 +\frac{1}{2}m_{\phi}^2\phi_0^2 +\frac{1}{4}d_3\phi_0^4 +\frac{1}{2}m_{\rho}^2\rho_{03}^2 \nonumber \\ &+&\Lambda_{\omega \rho}(g_{\omega N}^2\omega_0^2)(g_{\rho N}^2{\rho_{03}^2}) +\Lambda_{\phi \rho}(g_{\phi N}^2\omega_0^2)(g_{\rho N}^2{\rho_{03}^2}) \nonumber \\ 
&+& \sum_{B=N,Y} \frac{\gamma_B}{6\pi^2} \int_0^{{k_F}_B} \frac{k_B^4 dk}{\sqrt{k_B^2 + {m_B^{\star}}^2}} + \sum_{l=e,\mu} \frac{\gamma_l}{6 \pi^2} \int_0^{k_l} \frac{k_l^4 dk_l}{\sqrt{k_l^2 + m_l^2}} \nonumber \\ &-&\frac{1}{2}m_{\eta}^2\eta_0^2 +\frac{1}{2}m_{\xi}^2\xi_0^2 +\frac{\gamma_\chi}{6 \pi^2} \int_0^{k_F^{\chi}} \frac{k_\chi^4 dk_\chi}{\sqrt{k_\chi^2 + {m_\chi^{\star}}^2}} .
\label{pNL}
\end{eqnarray} 
Following our previous works \cite{Sen:2021wev, Guha:2021njn, Sen:2022pfr, Guha:2024pnn}, in the present work the values of $m_{\chi}$, $m_{\eta}$ and $m_{\xi}$ are consistently related with the self-interaction constraints from bullet cluster \cite{Randall:2007ph, Bradac:2006er, Tulin:2013teo, Tulin:2017ara, Hambye:2019tjt} while the self-interaction couplings $y_{\eta}$ and $y_{\xi}$ are also chosen by reproducing the observed non-baryonic relic density \cite{Belanger:2013oya, Gondolo:1990dk, Guha:2018mli}. The permitted values of $m_{\eta}$ and $m_{\xi}$ corresponding to the range of $m_{\chi}$ are already shown in our previous works \cite{Guha:2021njn, Sen:2021wev, Sen:2022pfr}. The direct detection, indirect detection, and collider search avenues for DM have excluded a vast region of the DM-SM interaction parameter space (SM stands for the standard model particles). Of late, light DM of mass in the sub-GeV order, has garnered a lot of attention. Therefore, in this work, we particularly focus on the $m_{\chi} \leq$ 1 GeV domain to understand the possible existence of DMANSs. The values of $m_{\eta}$, $y_{\eta}$, $y_{\xi}$, and $m_{\xi}$ corresponding to a particular value of $m_{\chi}$, which are considered in this work to obtain the priors for Bayesian analysis, are given in the Results section.

Within this range of $m_{\chi}$, we also restrict our investigation to a certain value of $\alpha$ for which the DM fraction ($f$) does not exceed 10\% for each hadronic model. Here,
\begin{eqnarray}
f=\frac{M}{M_{DM}} 
\end{eqnarray}
where, $M_{DM}$ is the contribution of DM to the total gravitational mass of the DMANSs and is given as
\begin{eqnarray}
M_{DM}=\int_0^R(4\pi r^2 \rho_{\chi} m_{\chi}) (1-2 M/r)^{-1/2}~dr .
\end{eqnarray}

For the outer crust, we consider the Baym-Pethick-Sutherland (BPS) EoS while the inner crust is taken care of with the EoS of the non-uniform matter produced with TM1 parameterization considering the self-consistent Thomas-Fermi approximation \cite{Bao:2015cfa}. The presence of DM along the density profile of the DMANSs is considered as an exponentially increasing function given by Eq. (\ref{eq:rho_chi}). According to this form, the value of $\rho_{\chi}$ and thus the presence of DM is significantly reduced towards the crust (low density regime). Therefore, we assume that the presence of DM can be neglected in the crust of the DMANSs. 
\begin{figure*}[ht]
\centering
\subfigure[\ Dark matter particle fraction]{\includegraphics[width=0.48\linewidth,height=7cm]{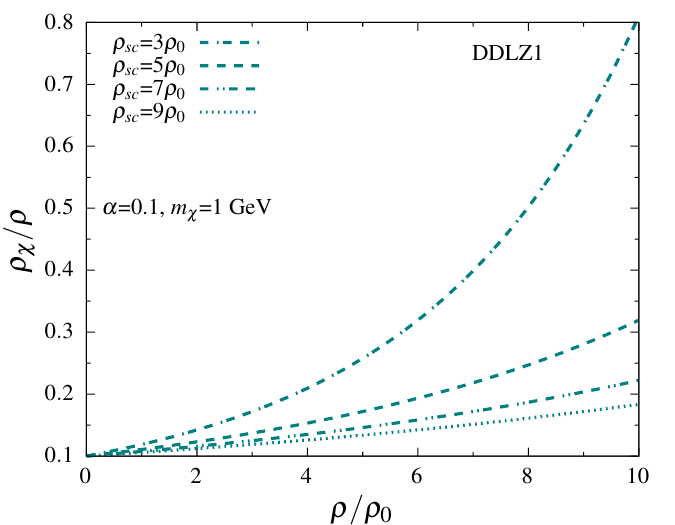} \label{fig:frac_rhosc_DDLZ1}}
\hfill 
\subfigure[\ Mass vs radius. The observational limits imposed on maximum mass from the most massive pulsar PSR J0740+6620 \cite{Fonseca:2021wxt} and corresponding radius \cite{Salmi:2024aum} are also indicated. The constraints on $M-R$ plane prescribed from GW170817 \cite{LIGOScientific:2018cki}, the NICER experiment for PSR J0030+0451 \cite{Vinciguerra:2023qxq}, HESS J1731-347 \cite{Doroshenko:2022}, PSR J0437-4715 \cite{Choudhury:2024xbk}, and PSR J1231-1411 \cite{Salmi:2024bss} are also compared.]{\includegraphics[width=0.48\linewidth,height=7cm]{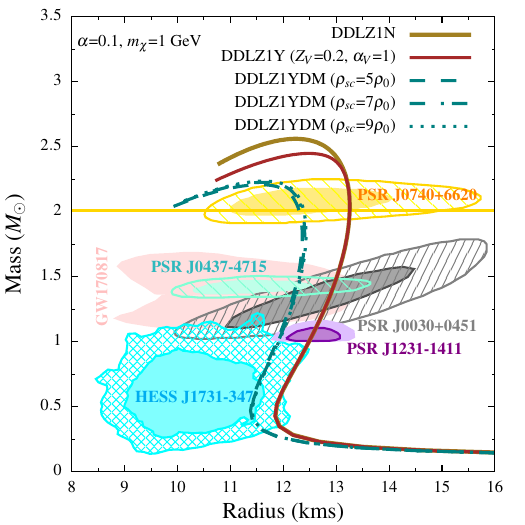}\label{fig:mr_rhosc_DDLZ1}}
\caption{(a) Dark matter particle fraction and (b) variation of mass with radius of dark matter admixed neutron stars for variation of $\rho_{sc}$ for fixed values of $\alpha$ and $m_{\chi}$ with DDLZ1 model.}
\label{fig:pf_mr_rhosc_DDLZ1}
\end{figure*}

With a sample values of $m_{\chi}$=1 GeV and $\alpha$=0.1, and for example, with the DDLZ1 model, in Fig. \ref{fig:pf_mr_rhosc_DDLZ1} we show the dependence of $\rho_{\chi}$ on the parameter $\rho_{sc}$ (Fig. \ref{fig:frac_rhosc_DDLZ1}) and its effect on the mass-radius relationship of the DMANSs (Fig. \ref{fig:mr_rhosc_DDLZ1}). We observe that the DM fraction increases with the decrease of $\rho_{sc}$ and for $\rho_{sc}$=3$\rho_0$, $\rho_{\chi}$ increases abruptly, following Eq. (\ref{eq:rho_chi}), and DM occupies almost 80\% of the NS matter, which is quite absurd in the case of DMANS scenario. Therefore, we study the variation of mass with radius without considering $\rho_{sc}$=3$\rho_0$. For other values of $\rho_{sc}$, we find that there is no significant change for the variation of $\rho_{sc}$. Therefore, in the present work, we fix a constant value of $\rho_{sc}$=5$\rho_0$ throughout our calculations. This reduces the number of free parameters of the DMANS to only two, viz. $m_{\chi}$ and $\alpha$. The values of $m_{\chi}$ and $\alpha$ of the present DMANS model are constrained by Bayesian analysis with respect to the observational constraints on the structural properties of compact stars. The method is discussed in the next section.

\section{Bayesian analysis of the parameters $\alpha$ and $m_{\chi}$ with respect to the observational constraints}
\label{Sec:Bayesian analysis}
The free parameters of the DMANSs can be constrained in the light of the existing observations regarding the structural properties of compact stars, e.g., the mass, radius and tidal deformability. For this purpose we take into account the following data sets which we include in our mass-radius diagram and tidal deformablity vs mass diagram. Simultaneous $M-R$ measurements of various pulsars (by NICER and HESS collaborations) and tidal deformability measurement of $1.4M_\odot$ compact star from GW170817 (by LIGO collaboration) are listed below.

\begin{enumerate}

\item  For PSR J0030+0451 $M = 1.40^{+0.13}_{-0.12}~M_\odot$ and $R = 11.71^{+0.88}_{-0.83}~\rm{km}$ reported by Vinciguerra {\it{et al.}}~\cite{Vinciguerra:2023qxq}.

\item  For PSR J0740+6620 $M = 2.08^{+0.07}_{-0.07}~M_\odot$ and $R = 12.92^{+2.09}_{-1.13}~\rm{km}$ reported by Dittmann {\it{et al.}}~\cite{Miller:2021qha}.

\item  For PSR J0437-4715 $M = 1.418^{+0.037}_{-0.037}~M_\odot$ and $R = 11.36^{+0.95}_{-0.63}~\rm{km}$ reported by Choudhury {\it{et al.}}~\cite{Choudhury:2024xbk}.

\item  For PSR J1231-1411 $M = 1.04^{+0.05}_{-0.03}~M_\odot$ and $R = 12.6^{+0.3}_{-0.3}~\rm{km}$ reported by Salmi {\it{et al.}}~\cite{Salmi:2024bss}.

\item  For PSR J0614-3329 $M = 1.44^{+0.06}_{-0.07}~M_\odot$ and $R = 10.29^{+1.01}_{-0.86}~\rm{km}$ reported by Mauviard {\it{et al.}}~\cite{Mauviard:2025dmd}.

\item For HESS J1731-347 $M = 0.77^{+0.20}_{-0.17}~M_\odot$ and $R = 10.04^{+0.86}_{-0.78}~\rm{km}$ reported by Doroshenko {\it{et al.}}~\cite{Doroshenko:2022}.

\item For GW170817 $\Lambda_{1.4} = 190^{+390}_{-120} $ reported by Abbott {\it{et al.}}~\citep{LIGOScientific:2018cki}.

\end{enumerate}
We infer the posterior distribution of the free parameters ($\alpha$ and $m_{\chi}$), defining the EoS, in the light of the observational data listed above. Consequently, we constrain the EoS to satisfy the observations. Bayes' theorem describes the posterior distribution of the parameters ($\alpha$ and $m_{\chi}$) for given observational data $\mathcal{D}$ as
\begin{eqnarray}
p\left((\alpha, m_{\chi}) \vert \mathcal{D} \right) = \frac{p\left( \mathcal{D} \vert (\alpha, m_{\chi})  \right) p(\alpha, m_{\chi})}{p(\mathcal{D})},
\end{eqnarray} 
where $p\left( \mathcal{D} \vert (\alpha, m_{\chi})  \right)$ is the full likelihood in terms of given observational data, $p(\alpha, m_{\chi})$ is the prior distribution of the parameters and $p(\mathcal{D})$ is the evidence. We follow \cite{Ayriyan:2024zfw} to estimate the full likelihood and the prior distribution is considered to be equiprobable within a range of reasonable guess values obtained from initial rough estimation. So,
\begin{eqnarray}
p(\alpha, m_{\chi}) = \frac{1}{\bigg\lvert \left(\alpha, m_{\chi}\right)\bigg\rvert} = \frac{1}{N}.
\end{eqnarray}
We calculate the likelihoods for all the independent observations ($\mathcal{D}_{j}$) and then the full likelihood is obtained by the overall product as
\begin{eqnarray}
p\left( \mathcal{D} \vert (\alpha, m_{\chi})  \right) = \prod_{j} p\left( \mathcal{D}_{j} \vert (\alpha, m_{\chi})  \right).
\end{eqnarray}
The evidence is estimated by summing all possible values of ($\alpha$ and $m_{\chi}$) taken into consideration; 
\begin{eqnarray}
p(\mathcal{D}) = \sum_{(\alpha, m_{\chi})} p\left( \mathcal{D} \vert (\alpha, m_{\chi})  \right) p(\alpha, m_{\chi}).
\end{eqnarray}
We compute the likelihood for each of the mass-radius constraints related to the observational pulsar data by integrating over the central energy density $\varepsilon_c$ with the appropriate probability density function \cite{Ayriyan:2024zfw},
\begin{eqnarray}
p\left( \mathcal{D}_{MR^{(i)}} \vert (\alpha, m_{\chi})  \right) = \int_{\varepsilon_c^{min}}^{\varepsilon_c^{max}(\alpha, m_{\chi})} f_{MR}^{(i)}  \left( M(\epsilon_c ; \alpha, m_{\chi}), R(M) \right) ~pr(\alpha, m_{\chi})~d \epsilon_c ,
\end{eqnarray}
where $ pr(\alpha, m_{\chi}) = 1/(\varepsilon_c^{max}(\alpha, m_{\chi}) - \varepsilon_c^{min})$. Following~\cite{Ayriyan:2024zfw}, we construct the probability density functions $f_{MR}^{(i)}$ using Kernel Density Estimation (KDE)~\cite{ChaconDuong:2018} on the basis of the data acquired from the Zenodo repository for the pulsars ($M-R$ constraints) \cite{zenodo}. For GW data, the likelihood is estimated in a similar fashion as an integration over $\varepsilon_c$,
\begin{eqnarray}
p\left( \mathcal{D}_{GW} \vert (\alpha, m_{\chi})  \right) = \int_{\varepsilon_c^{min}}^{\varepsilon_c^{max}(\alpha, m_{\chi})} f_{GW} \left( \Lambda_1(\epsilon_c ; \alpha, m_{\chi}), \Lambda_2(\Lambda_1) \right) ~pr(\alpha, m_{\chi})~d \epsilon_c,
\end{eqnarray}
where the probability density function $f_{GW}$ is built using the KDE based on the GW170817 data \cite{ligoLIGOP1800115v12GW170817}.

\section{Non-radial oscillation properties of dark matter admixed neutron stars}
\label{Sec:oscillation}

We calculate the fundamental ($f$-mode) and first pressure ($p_1$-mode) modes of the non-radial oscillation frequency of the DMANSs by adopting the full general relativistic formalism. The estimation of the oscillation frequency in a complete general relativistic approach is done by including both the perturbation of the matter inside the star and that of spacetime metric due to the non-radial oscillation \cite{1967ApJ...149..591T,Lindblom:1983ps,Lu:2011zzd,Thakur:2024ejl,Thorne:1967apj, Thorne:1969rba, Guha:2025ssq, Sen:2025ndg}. The complex eigen frequency can be written as $\omega=2\pi f + i/\tau$, where $\tau$ is the damping time of the GW. 

\section{Results}
\label{Results}
Focusing on the light ($m_{\chi}\leq$ 1 GeV) DM scenario, we initially consider three values of $m_{\chi}$=0.5, 0.75, 1 GeV, and for each of the three values of $m_{\chi}$, the three different values of $\alpha$ are taken to be 0.01, 0.05 and 0.1. These values of $m_{\chi}$ and $\alpha$ act as priors for the Bayesian analysis to constrain the free parameters (viz., $\alpha$ and $m_{\chi}$) of the DMANS model, in the light of recent observations. We tabulate below the DM masses and couplings corresponding to the chosen values of $m_{\chi}$ in Tab. \ref{tab:table_DM}.
\begin{table}[ht!]
\caption{Chosen values of $m_\chi$ and corresponding values of $m_\eta$ and $m_\xi$ from the constraints obtained from Bullet cluster. $y_\eta$ and $y_\xi$ have been fixed from observed relic abundance.}
{{
\setlength{\tabcolsep}{30.5pt}
\begin{center}
\begin{tabular}{ c c c c c c c c}
\hline
\hline
$m_{\chi}$ & $m_{\eta}$ & $m_{\xi}$ & $y_{\eta}$ & $y_{\xi}$ \\
(GeV) &(MeV) &(MeV)  &  \\
\hline
\hline
$0.5$ & $3.002$ & $3.002$ & $0.047$ & $0.022$ \\
$0.75$ & $3.5155$ & $4.00925$ & $0.059$ & $0.0203$ \\  
$1$ & $3.95$ & $4.9375$ & $0.07$ & $0.07$ \\  
\hline
\hline
\end{tabular}
\end{center}
}}
\protect\label{tab:table_DM}
\end{table}

We study the structural properties of the DMANSs with 15 hadronic models, viz., the BigApple ($L_0$=39.74 MeV), DDLZ1 ($L_0$=42.47 MeV), DDMEX ($L_0$=46.70 MeV), DDME2 ($L_0$=51.26 MeV), DD2 ($L_0$=54.95 MeV), DDME1 ($L_0$=55.46 MeV), NL3$\omega \rho 6$ ($L_0$=58 MeV), NL3$\omega \rho 5$ ($L_0$=61 MeV), NL3$\omega \rho 4$ ($L_0$=68 MeV), NL3$\omega \rho 3$ ($L_0$=77 MeV), NL3$\omega \rho 2$ ($L_0$=88 MeV), PKDD ($L_0$=90.12 MeV), GM1 ($L_0$=94.0 MeV), NL3$\omega \rho 1$ ($L_0$=101 MeV), and NL3 ($L_0$=118.32 MeV), with increasing values of symmetry energy and its slope parameter $L_0$ at saturation density $\rho_0$. We show the results of the structural properties of the DMANSs under static and rotating conditions and compare our results with the relevant astrophysical constraints explicitly for the BigApple, DD2, NL3$\omega \rho 6$, and GM1 models which have the lowest, moderate and high values of $L_0$, respectively, among the 15 chosen models. With the other models, we specify whether they can support the possible existence of light DM in DMANSs or not. The results with the DDME1 model along with those obtained with the BigApple model are shown later with Bayesian analysis.
\subsection{Structural properties}
\label{sec:Result_Structure}
The structural properties of the NSs in the presence and absence of hyperons are computed first without considering the DM. The two cases are denoted by `Y' and `N', respectively. Next, in the presence of hyperons, the contribution of DM is included for different values of $\alpha$ and $m_{\chi}$ to obtain the structural properties of the DMANSs under both static and rotational conditions.
\subsubsection{With BigApple model}
\label{subsec:BigAppleresults}
\begin{figure*}[ht]
\centering
\subfigure[\ Mass vs radius in static condition]{\includegraphics[width=0.32\textwidth]{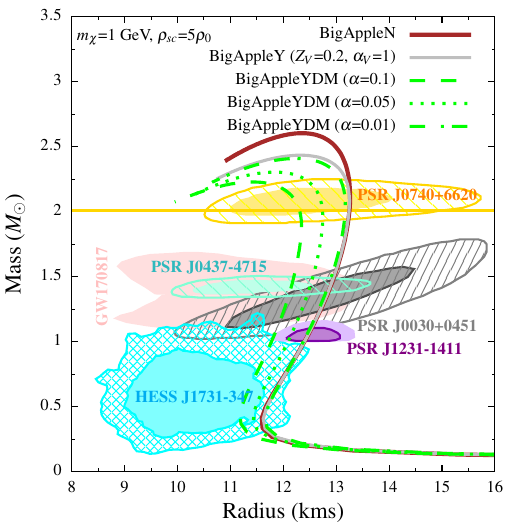} \label{fig:mr_alpha_BigApple}}
\hfill 
\subfigure[\ Corresponding tidal deformability vs mass. The constraints on $\Lambda_{1.4}$ from GW170817 \cite{LIGOScientific:2018cki} is also shown.]{\includegraphics[width=0.32\textwidth]{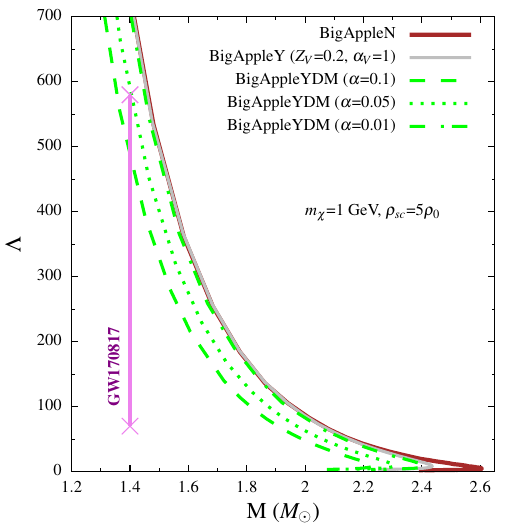}\label{fig:LamM_alpha_BigApple}}
\hfill
\subfigure[\ Mass vs radius in rotating condition. The constraint on rotational maximum mass ($M^R$=2.35 $\pm$ 0.17 $M_{\odot}$) \cite{Romani:2022jhd} from PSR J0952-0607 rotating at frequency $\nu$=707 Hz \cite{Bassa:2017zpe} is also compared.]{\includegraphics[width=0.32\textwidth]{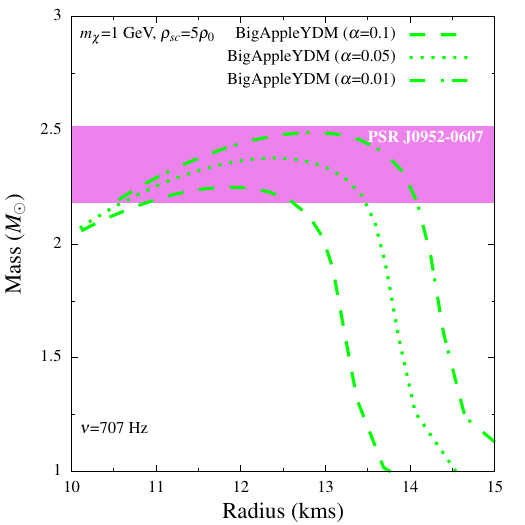} \label{fig:mr_rot_alpha_BigApple}}
\caption{Variation of (a) mass with radius in static condition, (b) corresponding tidal deformability with mass and (c) mass with radius in rotating condition (rotational frequency $\nu$=707 Hz) of dark matter admixed neutron stars for variation of $\alpha$ and fixed values of $\rho_{sc}$ and $m_{\chi}$ with BigApple model.}
\label{fig:MR_LamM_mr_rot_alpha_BigApple}
\end{figure*}
We start with the BigApple model, which has the minimum value of $L_0$ among all the hadronic models chosen for the present work. In Fig. \ref{fig:MR_LamM_mr_rot_alpha_BigApple} we first examine the impact of the free parameter $\alpha$ on the structural properties of the DMANSs for a fixed value of $m_{\chi}$=1 GeV. Before studying the DMANSs, we can see from Fig. \ref{fig:mr_alpha_BigApple} that the presence of hyperons (`Y') reduces the maximum mass to 2.43 $M_{\odot}$ from 2.60 $M_{\odot}$, the case where only nucleons (`N') are considered, without bringing any significant change to the corresponding radius. The onset of hyperons occurs at high density, and therefore the branching of stars with hyperons (`Y') from those without hyperons (`N') happens at high mass ($>$ 2 $M_{\odot}$). Both scenarios (`N' and `Y') satisfy all the constraints on the $M-R$ diagram of compact stars, like those obtained from pulsar observations of PSR J0740+6620 \cite{Fonseca:2021wxt,Miller:2021qha, Riley:2021pdl}, PSR J0030+0451 \cite{Riley:2019yda, Miller:2019cac}, HESS J1731-347 \cite{Doroshenko:2022}, PSR J0437-4715 \cite{Choudhury:2024xbk}, and PSR J1231-1411 \cite{Salmi:2024bss} and the GW170817 data \cite{LIGOScientific:2018cki}. However, Fig. \ref{fig:LamM_alpha_BigApple} shows that the values of tidal deformability ($\Lambda_{1.4}$) of a 1.4 $M_{\odot}$ mass star, obtained in both the presence (case `Y') and the absence (case `N') of hyperons, are inconsistent with the observational results of GW170817 \cite{LIGOScientific:2018cki}. On the other hand, we obtain DMANS configurations that can successfully satisfy the constraints on the mass, radius, and $\Lambda_{1.4}$ of compact stars. It can be seen that larger values of $\alpha$ make the DMANS configurations deviate more from the no-DM scenarios (the `N' and `Y' cases), as we observe that there is a progressive decrease in the maximum mass and radius of the DMANSs with the gradual increase in the value of $\alpha$. This is because this parameter $\alpha$ dictates the distribution of DM within the star (Eq. \ref{eq:rho_chi}) and higher values of $\alpha$ indicate a moderately high fraction of DM that can bring significant changes to the structural properties of DMANSs compared to those of the cases without DM. Thus, for a very low value of $\alpha$=0.01, the configuration almost overlaps with the scenario of NSs in the presence of hyperons (`Y'). Moreover, $\alpha$ cannot have very high values ($>$ 0.1) because such values do not satisfy the mass-radius constraint from PSR J1231-1411. In the present work we do not consider the value of $\alpha$ to be more than 0.1 to maintain the DM fraction ($f$) to a maximum of $\approx$ 10\% for chosen value of $\rho_{sc}$. It is also important to note that unlike NSs, DMANSs, with values of $\alpha<$ 0.05, do not have reasonable values of $\Lambda_{1.4}$ in terms of the GW170817 data. Therefore, we obtain an approximate range of $\alpha$ as 0.05--0.1, which can serve as prior for Bayesian analysis. In Fig. \ref{fig:mr_rot_alpha_BigApple} we also depict the variation of mass and radius of DMANSs rotating at frequency $\nu$=707 Hz, which happens to be the rotational frequency \cite{Bassa:2017zpe} of the most massive pulsar PSR J0952-0607 \cite{Romani:2022jhd} detected till date. The maximum masses of all the DMANSs, rotating at the frequency of PSR J0952-0607, are in accordance with the measured mass of this pulsar. We also notice that there is moderate increase in mass of the DMANSs in the rotational scenario compared to that in the static case but the radius of low mass stars is greatly affected due to such fast rotation.
\begin{figure*}[ht]
\centering
\subfigure[\ Mass vs radius in static condition]{\includegraphics[width=0.32\textwidth]{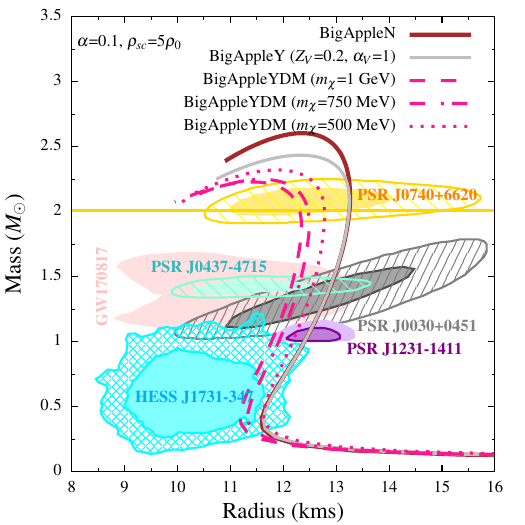} \label{fig:mr_mchi_BigApple}}
\hfill 
\subfigure[\ Corresponding tidal deformability vs mass]{\includegraphics[width=0.32\textwidth]{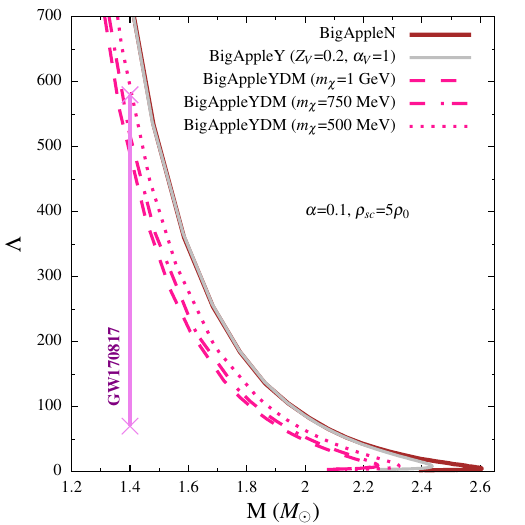}\label{fig:LamM_mchi_BigApple}}
\hfill 
\subfigure[\ Mass vs radius in rotating condition]{\includegraphics[width=0.32\textwidth]{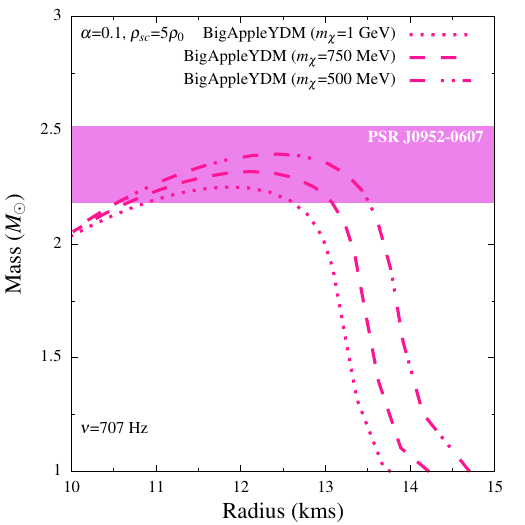} \label{fig:mr_rot_mchi_BigApple}}
\caption{Variation of (a) mass with radius in static condition, (b) corresponding tidal deformability with mass and (c) mass with radius in rotating condition (rotational frequency $\nu$=707 Hz) of dark matter admixed neutron stars for variation of $m_{\chi}$ and fixed values of $\alpha$ and $\rho_{sc}$ with BigApple model.}
\label{fig:MR_LamM_mr_rot_mchi_BigApple}
\end{figure*}

Next, we concentrate on the other free parameter $m_{\chi}$ to test its influence on the structure of DMANSs in Fig. \ref{fig:MR_LamM_mr_rot_mchi_BigApple} for a fixed value of $\alpha$=0.1. It is evident from Figs. \ref{fig:mr_mchi_BigApple} and \ref{fig:LamM_mchi_BigApple} that our results of the mass, radius and $\Lambda_{1.4}$ of DMANSs are in excellent accordance with all the astrophysical and observational constraints. Lower values of $m_{\chi}$ produce more massive DMANSs with a larger radius. However, all DMANSs are less massive and have a lesser radius compared to NSs. It is interesting to note that for $m_{\chi}>$ 1 GeV the constraint on $M-R$ plane from PSR J1231-1411 is not satisfied. This sets an upper limit on $m_{\chi}$ and also favors the presence of sub-GeV, light DM in DMANSs. From the variation of tidal deformability of DMANSs in Fig. \ref{fig:LamM_mchi_BigApple}, we notice that $m_{\chi}<$ 500 MeV yields values of $\Lambda_{1.4}$ that are inconsistent with GW170817 data. Therefore, the overall range of $m_{\chi}$ can be taken as (0.5--1) GeV as a prior for Bayesian analysis. The structural properties of DMANSs are more sensitive to the parameter $\alpha$ than $m_{\chi}$ because a slight variation of the former shows more significant changes than the latter. From Fig. \ref{fig:mr_rot_mchi_BigApple} it can be understood that the DMANS configurations obtained with different values of $m_{\chi}$ can successfully explain the possibility of massive and rapidly rotating pulsars like PSR J0952-0607 being DMANSs. 

\subsubsection{With DD2 model}
\label{subsec:DD2results}
\begin{figure*}[ht]
\centering
\subfigure[\ Mass vs radius in static condition]{\includegraphics[width=0.32\textwidth]{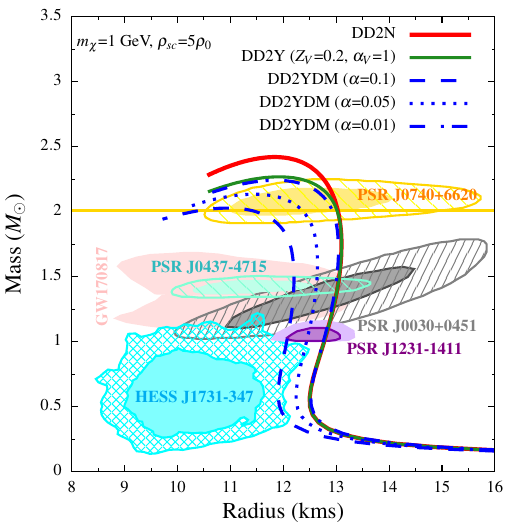} \label{fig:mr_alpha_DD2}}
\hfill 
\subfigure[\ Corresponding tidal deformability vs mass]{\includegraphics[width=0.32\textwidth]{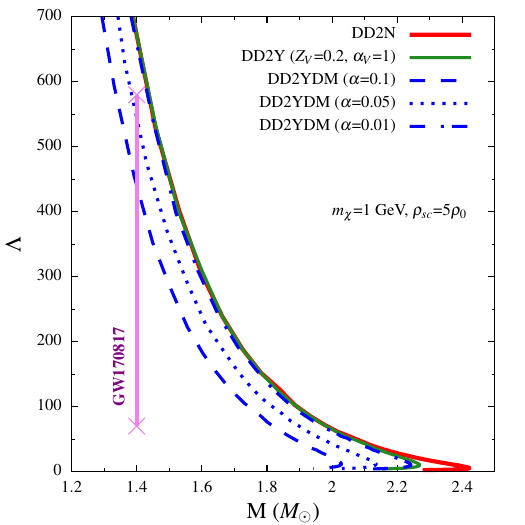}\label{fig:LamM_alpha_DD2}}
\hfill
\subfigure[\ Mass vs radius in rotating condition]{\includegraphics[width=0.32\textwidth]{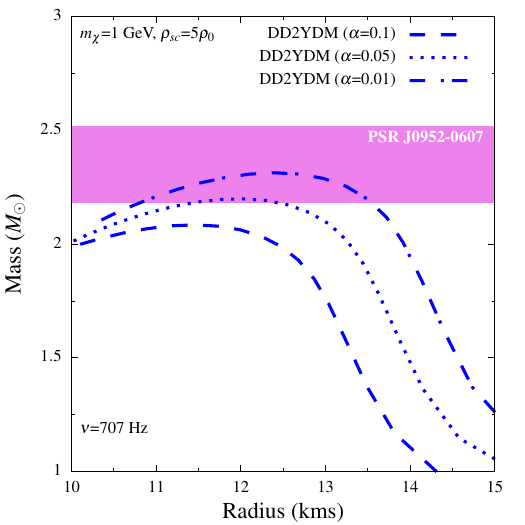} \label{fig:mr_rot_alpha_DD2}}
\caption{Variation of (a) mass with radius in static condition, (b) corresponding tidal deformability with mass and (c) mass with radius in rotating condition (rotational frequency $\nu$=707 Hz) of dark matter admixed neutron stars for variation of $\alpha$ and fixed values of $\rho_{sc}$ and $m_{\chi}$ with DD2 model.}
\label{fig:MR_LamM_mr_rot_alpha_DD2}
\end{figure*}
Subsequently, we consider the DD2 model which has an intermediate value of $L_0$=54.95 MeV among all the hadronic models chosen for the present work. It can be seen from Fig. \ref{fig:mr_alpha_DD2} that similar to the BigApple model, the maximum mass of NS is 2.42 $M_{\odot}$ in the absence of hyperons while it is 2.27 $M_{\odot}$ in its presence in the case of the DD2 model. However, unlike the BigApple model, the constraint from the pulsar HESS J1731-347 is satisfied by neither of the `N' or `Y' cases with the DD2 model. As expected, even DMANSs with very low values of $\alpha$=0.01 cannot be helpful to satisfy this constraint, although all the other constraints on the $M-R$ plot are harmoniously satisfied. Fig. \ref{fig:LamM_alpha_DD2} also confirms that the values of $\Lambda_{1.4}$, obtained with `N' or `Y' cases and DMANSs with $\alpha$ as low as 0.01, are not in par with the GW170817 data. Consistent with the analysis done for the BigApple model, we find that all the observational constraints on the mass, radius and tidal deformability are satisfied by the DMANSs only for 0.05 $\leq\alpha\leq$ 0.1 for fixed $m_{\chi}$=1 GeV. From Fig. \ref{fig:mr_rot_alpha_DD2} we find that the mass of PSR J0952-0607 is satisfied by rotating DMANSs with the frequency of the pulsar when we take $\alpha$=0.01--0.05. 
\begin{figure*}[ht]
\centering
\subfigure[\ Mass vs radius in static condition]{\includegraphics[width=0.32\textwidth]{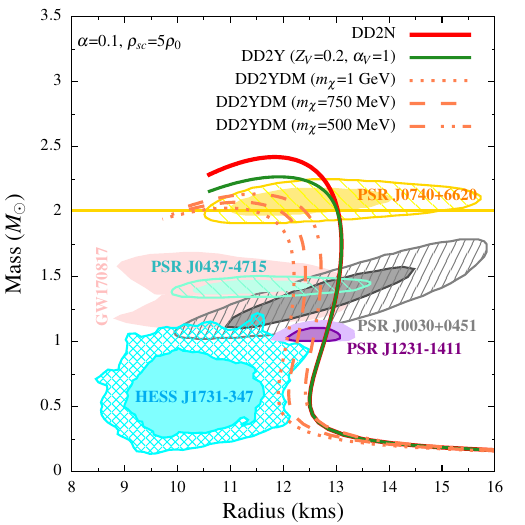} \label{fig:mr_mchi_DD2}}
\hfill 
\subfigure[\ Corresponding tidal deformability vs mass]{\includegraphics[width=0.32\textwidth]{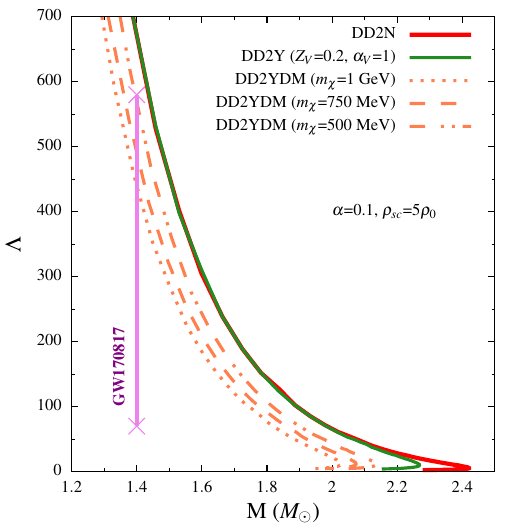}\label{fig:LamM_mchi_DD2}}
\hfill
\subfigure[\ Mass vs radius in rotating condition]{\includegraphics[width=0.32\textwidth]{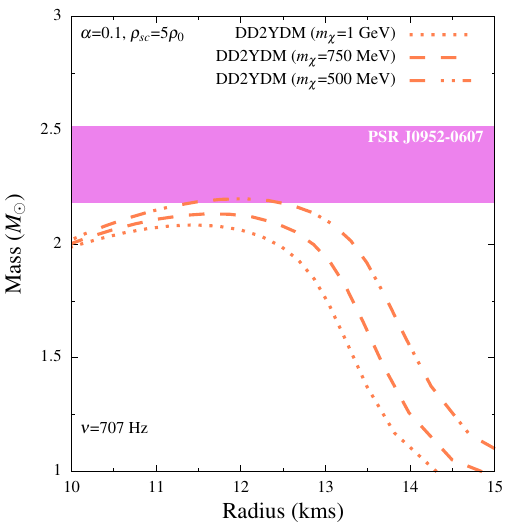} \label{fig:mr_rot_mchi_DD2}}
\caption{Variation of (a) mass with radius in static condition, (b) corresponding tidal deformability with mass and (c) mass with radius in rotating condition (rotational frequency $\nu$=707 Hz) of dark matter admixed neutron stars for variation of $m_{\chi}$ and fixed values of $\alpha$ and $\rho_{sc}$ with DD2 model.}
\label{fig:MR_LamM_mchi_DD2}
\end{figure*}

By fixing $\alpha$=0.1, we now seek the reasonable range of $m_{\chi}$ for DD2 model in the light of the different observational constraints. From Figs. \ref{fig:mr_mchi_DD2} and \ref{fig:LamM_mchi_DD2} we conclude that the range of $m_{\chi}$=(0.5--1) GeV, obtained for the BigApple model, also serves as an excellent choice for the DD2 model. It can be easily verified that DMANS configurations with $m_{\chi}<$ 500 MeV will not be compatible with the mass-radius data of HESS J1731-347 and $\Lambda_{1.4}$ data of GW170817 while for $m_{\chi}>$ 1 GeV the mass-radius data of PSR J1231-1411 will be violated. It is also seen in Fig. \ref{fig:mr_rot_mchi_DD2} that the mass of the PSR J0952-0607 is satisfied with its rotational frequency by the DMANS having $m_{\chi}$=500 MeV with the DD2 model.

For the DDME1 model, the value of $L_0$ is 55.46 MeV, which is very close to that of the DD2 model. Therefore, we obtain almost similar results for the DDME1 model as for the DD2 model. For the same range of $\alpha$=0.01-0.1 and $m_{\chi}$=(0.5-1) GeV, the structural properties of DMANSs with the DDME1 model satisfy all the astrophysical constraints under both static and rotating conditions. The results of the initial analysis of the BigApple, DD2, and DDME1 models jointly confirm that to obtain reasonable DMANS configurations, the suitable ranges of the two free parameters $\alpha$ and $m_{\chi}$ do not depend on the value of $L_0$ or the underlying hadronic model within a range of $L_0$=(39.74-55.46) MeV. Within this range, irrespective of the properties of the hadronic model, the presence of light DM (of mass in sub-GeV order) in moderate amounts ($f <$ 10\%) can support the possible existence of DMANSs.

\begin{figure*}[t]
\centering
\subfigure[\ Variation of $\alpha$]{\includegraphics[width=0.35\textwidth]{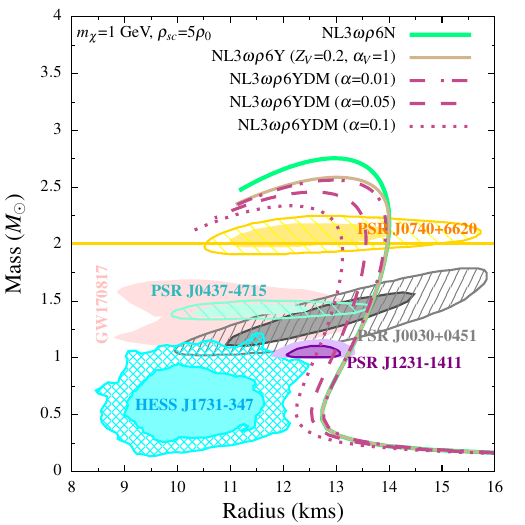} \label{fig:mr_alpha_NL3wr6}}
\subfigure[\ Variation of $m_{\chi}$]{\includegraphics[width=0.35\textwidth]{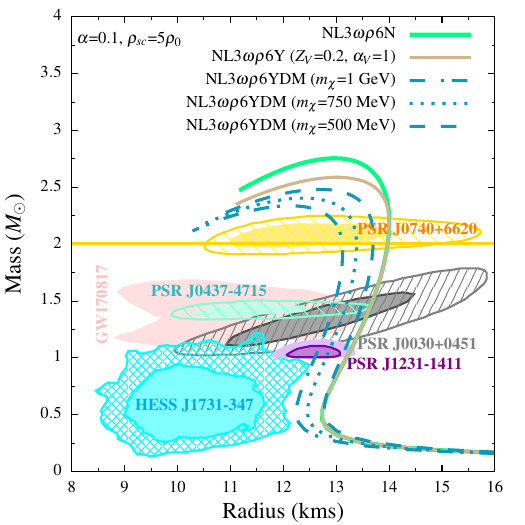} \label{fig:mr_mchi_NL3wr6}}
\caption{Dependence of mass on radius in static condition of dark matter admixed neutron stars for variation of (a) $\alpha$ and (b) $m_{\chi}$ with fixed values $\rho_{sc}$ with NL3$\omega \rho 6$ model.}
\label{fig:MR_alpha_mchi_NL3wr6}
\subfigure[\ Variation of $\alpha$]
{\includegraphics[width=0.35\textwidth]{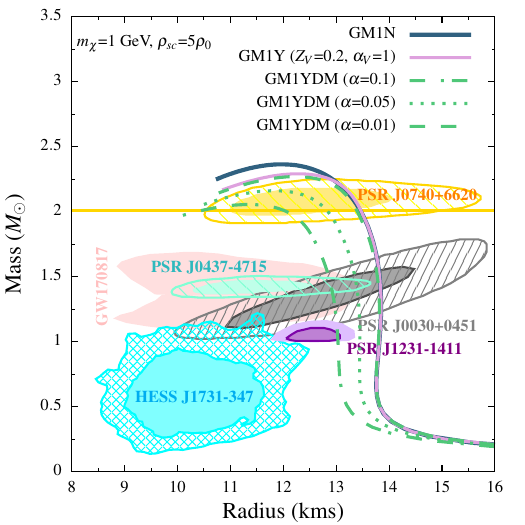} \label{fig:mr_alpha_GM1}}
\subfigure[\ Variation of $m_{\chi}$]{\includegraphics[width=0.35\textwidth]{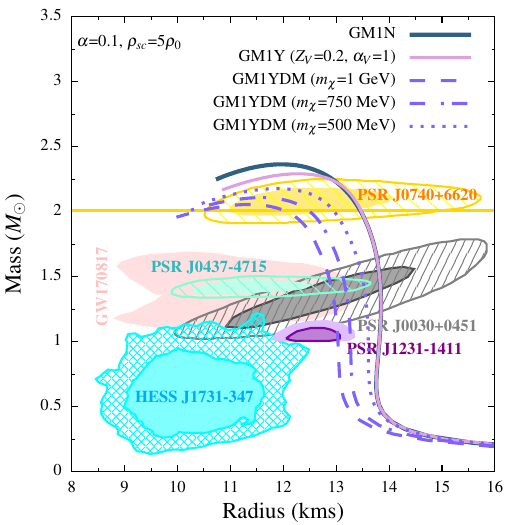} \label{fig:mr_mchi_GM1}}
\caption{Dependence of mass on radius in static condition of dark matter admixed neutron stars for variation of (a) $\alpha$ and (b) $m_{\chi}$ with fixed values $\rho_{sc}$ with GM1 model.}
\label{fig:MR_alpha_mchi_GM1}
\end{figure*}

\subsubsection{With NL3$\omega \rho 6$ and GM1 models}
\label{subsec:NL3omegarho5_GM1results}

Finally, we examine the structural properties of the DMANSs with the NL3$\omega \rho 6$ and GM1 models. Ref. \cite{Lattimer:2023rpe} has shown that the constraints from PREX and CREX experiments, GW170817 and other astrophysical observations, are satisfied within the range $L_0=53 \pm 13$ MeV. The lower limit of $L_0$ from this range is considered in the present work with the BigApple model. The NL3$\omega \rho 6$ model is taken into account, for which the value of $L_0$(=58 MeV) is close to the upper limit suggested by \cite{Lattimer:2023rpe}. The mass-radius relationship of DMANSs is studied with the NL3$\omega \rho 6$ model for the variation of $\alpha$ and $m_{\chi}$ in Figs. (\ref{fig:mr_alpha_NL3wr6}) and (\ref{fig:mr_mchi_NL3wr6}), respectively. The same is done with the GM1 model, having a very large value of $L_0$=94 MeV in Figs. (\ref{fig:mr_alpha_GM1}) and \ref{fig:mr_mchi_GM1}. It can be comprehended from Figs. \ref{fig:mr_alpha_NL3wr6} and \ref{fig:mr_alpha_GM1}, that NS configurations with the NL3$\omega \rho 6$ model in the presence and absence of hyperons fail to satisfy several important astrophysical constraints like the $M-R$ from HESS J1731-347, and GW170817. With the GM1 model, these constraints along with that from PSR J1231-1411 are all unfulfilled in the presence and absence of hyperons. We note that for such models with very large values of $L_0$, even the presence of light DM in moderate amount ($f \leq$ 10\%) cannot fulfill the HESS J1731-347 data although all the other observational data are in good agreement with the results of DMANSs obtained with the same range of $\alpha$ and $m_{\chi}$ inferred from the analysis of the BigApple and DD2 models. It can be noticed from the Fig. \ref{fig:mr_alpha_GM1} that considering values of $\alpha$ and $m_{\chi}$ even beyond the range obtained from the BigApple and DD2 models will not satisfy all the observational data simultaneously with models having very large values of $L_0$. For example, for the NL3$\omega \rho 6$ model, any value of $\alpha >$ 0.1 will surely satisfy the HESS J1731-347 data but in such cases $f > 10\%$. Similarly, for the GM1 model, considering $m_{\chi} >$ 1 GeV will meet the constraint from HESS J1731-347 but the maximum mass will be too small to satisfy the maximum mass constraint from PSR J0740+6620. 

Considering all the hadronic models, we emphasize here the importance of a certain upper limit of $L_0$ and the HESS J1731-347 data to determine the possible existence of DMANSs with light DM. The constraint on $\Lambda_{1.4}$ from GW170817 is also equally important in this context because within this range of $L_0$, this constraint is satisfied only in presence of DM. In Tab. \ref{tab:L} we provide whether the models with different values of $L_0$ for DMANSs in the presence of hyperons (NSYDM) can satisfy all the astrophysical constraints or not. It is clear from Tab. \ref{tab:L} that $L_0$ can play a crucial role in determining the existence of DMANSs. Since the exact value of $L_0$ is still inconclusive, and CREX and PREX experiments together with GW170817 data often emphasize lower values of $L_0$, we can consistently infer that smaller values of $L_0$ is preferred to support the possible existence of DMANSs. Among the chosen models starting from BigApple in ascending values of $L_0$, DDME1 is the last model to support the presence of DMANSs. It is also interesting to note that over this wide range of $L_0$=(39.74--55.46) MeV, DMANSs can have a very tight range of $\alpha$ and $m_{\chi}$, indicating that the presence of light DM of mass in sub-GeV order can produce reasonable DMANSs when such type of DM populates the NS matter in a moderate amount (up to $f \leq$ 10\%). In case of the NL3$\omega \rho 6$ model, it can be understood from Fig. \ref{fig:MR_alpha_mchi_NL3wr6} that simultaneous fulfillment of all the constraints on the $M-R$ plane is possible for $\alpha>$ 0.1 and/or $m_{\chi}>$ 1 GeV. However, these values are out of the range obtained with the six models for a wide range of $L_0=39.74-55.46$ MeV. Model-independent ranges of $\alpha=0.01-0.1$ and $m_{\chi}=0.5-1$ GeV within an acceptable range of $L_0=53 \pm 13$ MeV, are obtained only up to $L_0=$ 55.46 MeV starting from $L_0=$ 39.74 MeV from the collective results of the six models viz., BigApple, DDLZ1, DDMEX, DDME2, DD2, and DDME1. Moreover, $\alpha>$ 0.1 implies the presence of high fraction of DM ($f$ more than 10\%) and $m_{\chi}>$ 1 GeV is also not favored by the various recent DM search experiments. Therefore, we infer that up to a maximum $L_0 \approx 56$ MeV, reasonable amount of light DM can be present in NSs to support the possible existence of DMANS in the light of the various astrophysical constraints.   
\begin{table}[!ht]
\caption{The models (with ascending values of $L_0$) for which the neutron stars in presence of only hyperons (NSY (no-DM)) and dark matter admixed neutron stars in presence of hyperons and light dark matter (NSYDM) satisfy (denoted by $"\checkmark"$) or do not satisfy (denoted by $"\times"$) all the astrophysical constraints (listed in Sec. \ref{Sec:Bayesian analysis}) on the structural properties of compact stars.}
\setlength{\tabcolsep}{7.0pt}
\begin{tabular}{c|c|c|c|c|c}
\hline
\hline
Model & $L_0$ (MeV) & \multicolumn{2}{c|}{NSY (no-DM)}  & \multicolumn{2}{c}{NSYDM} \\
\cline{3-6} 
& & $M-R$ (PSR Observations)  & $\Lambda_{1.4}$ (GW170817) & $M-R$ (PSR Observations) & $\Lambda_{1.4}$ (GW170817)\\
\hline
BigApple           & 39.7407  & $\checkmark$  & $\times$ & $\checkmark$  & $\checkmark$\\
DDLZ1              & 42.4660  & $\checkmark$    & $\times$ & $\checkmark$ & $\checkmark$\\
DDMEX              & 46.6998    & $\times$ & $\times$ & $\checkmark$ & $\checkmark$\\
DDME2              & 51.2653    & $\times$ & $\times$ & $\checkmark$ & $\checkmark$\\
DD2                & 54.9529    & $\times$ & $\times$ & $\checkmark$ & $\checkmark$\\
DDME1              & 55.4634    & $\times$ & $\times$ &  $\checkmark$ & $\checkmark$\\
NL3$\omega \rho 6$ & 58         & $\times$ & $\times$ & $\times$ & $\times$ \\
NL3$\omega \rho 5$ & 61         & $\times$ & $\times$ & $\times$ & $\times$\\ 
NL3$\omega \rho 4$ & 68         & $\times$ & $\times$ & $\times$ & $\times$\\ 
NL3$\omega \rho 3$ & 77         & $\times$ & $\times$ & $\times$ & $\times$\\
NL3$\omega \rho 2$ & 88         & $\times$ & $\times$ & $\times$ & $\times$\\
PKDD               & 90.1204    & $\times$ & $\times$ & $\times$ & $\times$\\
GM1                & 94         & $\times$ & $\times$ &  $\times$ & $\times$\\
NL3$\omega \rho 1$ & 101        & $\times$ & $\times$ & $\times$ & $\times$\\
NL3                & 118.3225   & $\times$ & $\times$ & $\times$ & $\times$\\
\hline
\hline
\end{tabular}
\label{tab:L}
\end{table}
\subsection{Bayesian analysis to constrain $\alpha$ and $m_{\chi}$}
\label{sec:Bayesian}
\begin{figure*}[ht]
\centering
\subfigure[~BigApple model]{\includegraphics[width=0.49\textwidth]{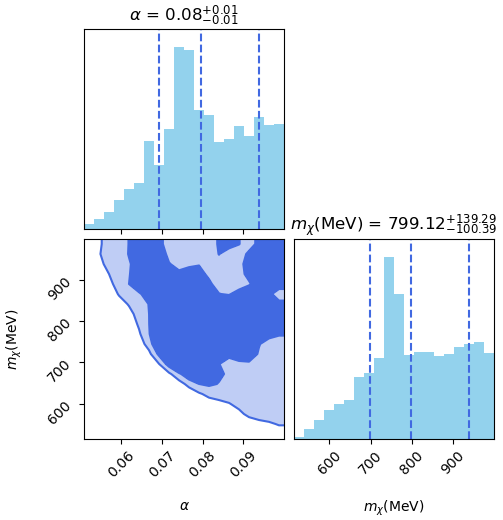} 
\label{fig:Bayesian_BigApple}}
\hfill
\subfigure[~DDME1 model]
{\includegraphics[width=0.49\textwidth]{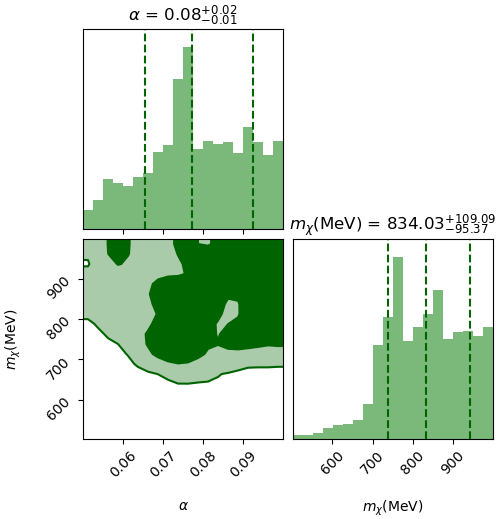} 
\label{fig:Bayesian_DDME1}}
\caption{Posterior distribution of the free parameters $\alpha$ and $m_{\chi}$ for (a) BigApple model and (b) DDME1 model obtained from Bayesian analysis with the 68\% and 95\% credible intervals.}
\end{figure*}
To obtain the specific ranges for the two free parameters of our DMANS model (viz. $\alpha$ and $m_{\chi}$) that are capable of producing reasonable DMANSs with respect to all the astrophysical constraints, we perform a Bayesian analysis separately for the BigApple and DDME1 models. We follow the formulation of the Bayesian analysis as discussed in Sec. \ref{Sec:Bayesian analysis}. The results of the previous Sec. \ref{sec:Result_Structure} guide us to establish proper prior distributions for the analysis as $\alpha$ = 0.01--0.1 and $m_{\chi}$ = 0.5--1 GeV. The posterior distributions of the parameters are obtained after performing the Bayesian analysis. We show the posterior distributions in terms of the $68\%$ and $95\%$ credible intervals in Figs. \ref{fig:Bayesian_BigApple} and \ref{fig:Bayesian_DDME1} for BigApple and DDME1 models, respectively. The most credible ranges (MCRs) and Maximum A-Posteriori (MAP) estimates for the two parameters are listed below in Tab. \ref{tab:MAP} for the two hadronic models.
\begin{table}[!ht]
\caption{The most credible ranges (MCRs) and Maximum A-Posteriori (MAP) estimates of the parameters $\alpha$ and $m_{\chi}$ obtained from Bayesian analysis for the BigApple and DDME1 models.}
\setlength{\tabcolsep}{25.0pt}
\begin{tabular}{ccccccc}
\hline
\hline
Model & MCR of $\alpha$ & MCR of $m_{\chi}$ (MeV) & MAP of $\alpha$ & MAP of $m_{\chi}$ (MeV) \\
\hline
BigApple     & 0.07--0.09  & 698.73--938.41 & 0.08 & 762.75 \\
DDME1        & 0.07--0.10  & 738.66--943.12 & 0.08 & 763.64 \\
\hline
\hline
\end{tabular}
\label{tab:MAP}
\end{table}
From Tab. \ref{tab:MAP} it is very clear that the MCRs of both $\alpha$ and $m_{\chi}$ are very close to each other for BigApple and DDME1 although the two hadronic models differ largely in their characteristics of symmetry energy and the value of $L_0$. It can be concluded that hadronic models, characterized by a low to moderate value of $L_0$, can support the existence of DMANSs. In addition, interestingly, our MAP estimates for the two models are enough to show that the amount of DM and its mass required by the DMANS to fulfill the observational constraints on compact star are independent of the underlying model and the value of $L_0$. Thus, the estimates of MAP, obtained for $\alpha$ and $m_{\chi}$, can be considered universal with respect to the hadronic model that favors the existence of DMANSs. The results also ensure that a suitable fraction ($f <$ 10\%) of light DM ($m_{\chi} <$ 1 GeV) is enough to produce reasonable DMANSs whose structures are compatible with the observational constraints on compact stars. For the BigApple and DDME1 models, the MAP values give $f_{MAP}$=9.49 \% and $f_{MAP}$=9.66 \%, respectively for $M=M_{max}$.

\subsection{Oscillation properties}
\label{sec:oscillation}
\begin{figure*}[ht]
\centering
\subfigure[~$f$-mode]{\includegraphics[width=0.49\textwidth]{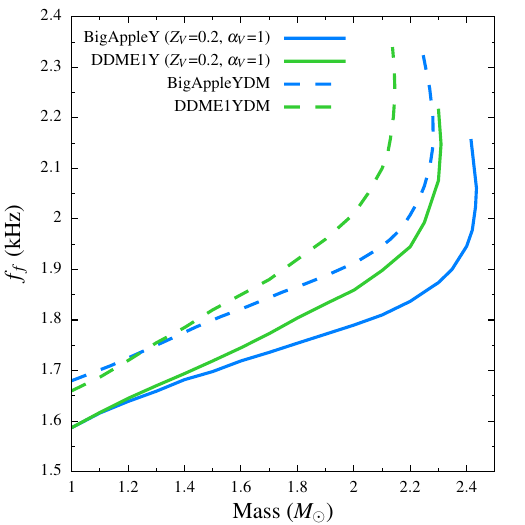} 
\label{fig:mf}}
\hfill
\subfigure[~$p_1$-mode]
{\includegraphics[width=0.49\textwidth]{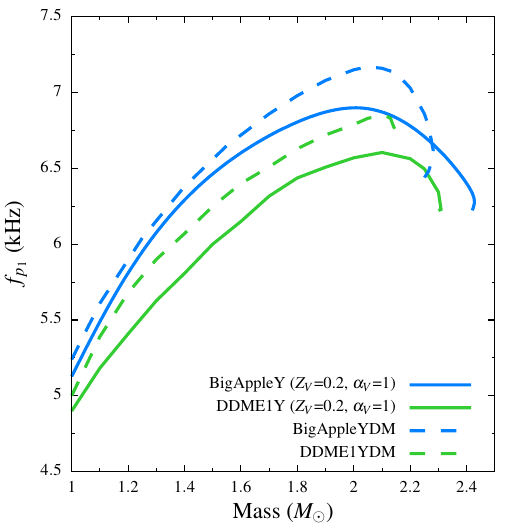} 
\label{fig:mp}}
\caption{Variation of (a) $f$-mode and (b) $p_1$-mode oscillation frequency with mass of dark matter admixed neutron stars for the BigApple and DDME1 models obtained with the Maximum A-Posteriori (MAP) values from Bayesian analysis.}
\end{figure*}
We calculate the $f$-mode and the $p_1$-mode frequencies, denoted by $f_f$ and $f_{p_1}$, respectively, in full GR treatment. For the BigApple and DDME1 models, the results are displayed in Figs. \ref{fig:mf} and \ref{fig:mp} showing the values of $f_f$ and$f_{p_1}$, respectively. In these figures, we study the effects of DM on $f_f$ and $f_{p_1}$ by comparing with the `no-DM' case in the presence of hyperons. We calculate $f_f$ and $f_{p_1}$ of the DMANSs with the MAP values obtained in Sec. \ref{sec:Bayesian} for the BigApple and DDME1 models. It can be seen from Fig. \ref{fig:mf} that $f_f$ increases in the presence of DM compared to the `no-DM' case. For example, in the BigApple model $f_{f_{1.4}}$=1.682 kHz in the absence of DM but for DMANS $f_{f_{1.4}}$=1.786 kHz.

Fig. \ref{fig:mp} also suggests that the presence of DM increases the value of $f_{p_1}$ with respect to the `no-DM' scenario. For example, in the DDME1 model $f_{{p_1}_{1.4}}$=5.807 kHz without DM but for DMANS $f_{{p_1}_{1.4}}$=6.07 kHz. The measurement of these oscillation frequencies by the upcoming GW detectors will enrich our understanding in this context. Further, simultaneous detection of tidal waves and non-radial modes of oscillation from BNS mergers will throw more light on constraining the EoS of DMANSs.


\section{Summary and Conclusion}
\label{Conclusion}

In this work, we explored the influence of DM in the NS, whose core is made up of nucleons and hyperons in the baryon octet. Several RMF models are chosen to describe the interactions of the nucleons and the hyperons. EoS of the NS matter becomes uncertain as the density increases. Uncertainty due to the symmetry energy is taken into account by considering the slope parameter $L_0$ in a wide range 40--120 MeV. 

At densities above the saturation density, chemical equilibration allows existence of exotic degrees of freedom. Among several possibilities, we consider the creation of hyperons via flavor changing weak-decay of the nucleons. We adopted the SU(3) coupling scheme to determine the coupling constants of the
hyperons with mesons. Parameters ($Z_V$ and $\alpha_V$) that are crucial to EoS in the presence of hyperons, are fixed by using the data of modern observation of heavy pulsars with mass $ > 2 M_\odot$.

Creation of hyperons in the NS softens the EoS, but in all the models we consider, the density at which hyperons populate is within 2.5--3 $\rho_0$, which is slightly delayed compared to those obtained with SU(6) coupling scheme \cite{Weissenborn:2011ut, Miyatsu:2013yta}. As a result, hyperons reside in stars whose mass is larger than $2 M_\odot$, so the observational maximum mass constraint from PSR J0740+6620 is always satisfied in the presence of hyperons. However, several important astrophysical constraints like the HESS J1731-347 data are not satisfied with most of the RMF models having $L_0 \gtrsim 43$ MeV. Also, with none of the hadronic models, the constraint on $\Lambda_{1.4}$ from GW170817 is satisfied.

In order to overcome the above issues, we invoke feeble interaction between hadronic matter and light fermionic DM $\chi$ via dark scalar and vector mediators, $\eta$ and $\xi$. $m_{\eta}$ and $m_{\xi}$ are related to $m_{\chi}$ using the self-interaction constraints from bullet cluster and the DM self-interaction couplings are related to $m_{\chi}$ through the relic density constraint. We consider the DM density as an exponential function of the baryon density with a free parameter $\alpha$. Certain experiments like LZ, XENON, DarkSide, CRESST, and LHC have almost ruled out the existence of massive DM in GeV order. Therefore, up to this upper limit of $m_{\chi}$, uncertainty from the dark sector is thoroughly explored via $m_{\chi}$ and $\alpha$ such that the contribution of DM to the total mass of the DMANSs is always $< 10\%$.

Inclusion of DM whose mass is below 1\,GeV also softens the EoS. Contrary to hyperons whose effect is limited to massive stars, DM also affects the bulk properties of the NSs in the region $M \gtrsim 0.5 M_\odot$. The effect is especially notable for the mass below $1.5 M_\odot$, in which region many new data from several pulsars are reported recently. Therefore, the DMANSs configurations obtained with light DM successfully satisfy the observational mass-radius data of low-mass pulsars like HESS J1731-347 as well as those of the high-mass pulsars like PSR J0740+6620 and intermediate-mass pulsars like PSR J0030+0451. Mass-radius behavior of the DMANSs is sensitive to the mass of DM and its density in the star. Using the various astronomical data from different pulsars and GW170817 as prior inputs, we applied Bayesian analysis to infer the posterior distributions of the parameters $m_{\chi}$ and $\alpha$. Results for both these parameters are weakly dependent on the hadronic model, and the standard deviation is about 12\,\% of the most probable value for both DM mass and DM density parameter.

Most probable values obtained from the Bayesian inference were plugged into the calculation of observables for the future measurement. We considered $l=2$ mode oscillations of the NSs and calculated $f$- and $p_1$-mode frequencies using BigApple and DDME1 models. Inclusion of light DM makes $f$-mode frequency increase by about 0.1--0.15\,kHz in the two models for the mass above $M_\odot$. $f_f$ is insensitive to the hadronic model at mass $\lesssim 1.4 M_\odot$, but dependence on the models becomes pronounced as the mass increases. In the $p_1$-mode oscillation, contribution of DM appears differently in the two models for the mass from $M_\odot$ to the maximum configuration. DDME1 model exhibits comparatively stronger dependence on the DM contribution than the BigApple model, especially in the low-mass regime ($< 1.4 M_\odot$).

Value of $L_0$ could also be constrained only in the presence of DM because without DM, no model can satisfy the constraint on $\Lambda_{1.4}$ from GW170817. With DM, we obtain a feasible range $40 \lesssim L_0 < 58$ MeV. Compared to the non-relativistic models like Skyrme force, RMF models tend to have stiffer EoSs. In a recent work \cite{ll2026}, nucleon-$\Lambda$ and $\Lambda$-$\Lambda$ interactions are determined by using single-$\Lambda$, double-$\Lambda$ hyper-nuclear data and observed NS properties with the KIDS (Korea:IBS-Daegu-SKKU) density functional. With the parameter sets obtained in \cite{ll2026}, one can extend the analysis of the present work to examine the dependence on the $\Lambda$-$\Lambda$ interactions as well as the symmetry energy in the presence of DM.



\section*{Acknowledgements}
Work of D.S. and C.H.H. is supported by the NRF research Grant (No. 2023R1A2C1003177). Work of A.G. is supported by the National Research Foundation of Korea (MSIT) (RS-2024-00356960).

\bibliography{ref}

\end{document}